\documentclass[11pt]{article}
\usepackage[a4paper, margin=3cm]{geometry}

\usepackage[utf8]{inputenc}
\usepackage[T1]{fontenc}

\usepackage{amsmath,amssymb,amsfonts,amsthm,mathrsfs}
\usepackage{bm}
\usepackage{siunitx}
\usepackage{graphicx}
\usepackage{caption}
\usepackage{booktabs}
\usepackage{multirow}
\usepackage{listings}
\usepackage{xcolor}
\usepackage{textcomp}
\usepackage{manyfoot}
\usepackage{authblk}
\usepackage[title]{appendix}
\usepackage[super,sort&compress,comma]{natbib}
\providecommand{\JournalTitle}[1]{#1}

\usepackage[
    colorlinks=true,
    linkcolor=blue,
    filecolor=blue,
    urlcolor=blue,
    citecolor=blue
]{hyperref}

\begin{document}

\setlength{\leftskip}{0pt}
\setlength{\rightskip}{0pt}
\setlength{\linewidth}{\textwidth}

\title{Digital twin for full field-of-view \\ Fraunhofer holographic patterning}
\author[1,*]{Antoni J. Wojcik} 
\author[1]{Dilawer Singh}
\author[1]{Hannah J. Joyce}
\author[1]{Timothy D. Wilkinson}
\affil[1]{Electrical Engineering Division, Department of Engineering,
  University of Cambridge, 9 JJ Thomson Avenue, Cambridge, CB3 0FA, UK}
\affil[*]{\rule{0pt}{3ex}Corresponding author: \texttt{ajw308@cam.ac.uk}}
\date{}
 
\maketitle 

\begin{abstract}
Computer-generated holography allows for nearly complete dynamic control of optical fields, but discrepancies between idealised propagation models and physical hardware limit its performance, particularly across large fields of view. We present a compact differentiable model of a Fraunhofer holographic projector which describes the complete propagation chain, from a phase-only spatial light modulator (SLM) to the camera recording the focal plane of a Fourier lens. It uses physically interpretable parameters for the SLM, illumination, pupil, coherent stray light, and camera. The model is identified in situ from noise-probed hologram-image pairs without additional hardware, system modification, or camera-in-the-loop refinement. The recovered digital twin is used to generate holograms for patterns spanning the entire first diffraction order of the SLM at its spatial-bandwidth limit. This full-field capability is enabled by measuring field-dependent pupil aberrations and compensating severe inter-pixel crosstalk by operating at the $4\pi$ phase modulation range. The stray light responsible for residual zeroth-order artefacts is recovered in amplitude and phase, allowing its physical origin to be investigated, and its contribution suppressed during hologram synthesis. The work establishes a framework for calibration-aware field synthesis in holographic systems operating in the Fraunhofer regime for applications such as maskless lithography and optical tweezer arrays.
\end{abstract}

\section{Introduction}\label{sec:intro}

Computer-generated holography (CGH)\cite{Osten:14, Javidi:21} shapes optical fields through controlled diffraction and interference of coherent light, with applications ranging from near-eye displays\cite{Chang:20} and optical trapping\cite{Grier:06, PhysRevX.4.021034, app122010244} to laser fabrication\cite{Bay:10, 4159977} and microscopy\cite{Park2018}. Hologram generation is an inverse problem, and algorithms for solving it have been studied extensively\cite{Pi2022, Sui2024}. However, their practical performance is limited by deviations between the idealised propagation model and the physical hardware, particularly in Fourier-domain systems, where such errors degrade image quality across the entire field of view (FoV). Accurate projection therefore requires accounting for system-specific imperfections.

System imperfections include optical misalignment\cite{Ruben:64}, non-uniform illumination\cite{7078844,Wojcik2025}, spatial light modulator (SLM) characteristics such as nonlinear phase response\cite{Fuentes:16, Xun:04} in liquid-crystal on silicon (LCoS) devices\cite{Zhang2014}, which also may be spatially variant\cite{Engstrom:13, Reichelt:13}, inter-pixel crosstalk\cite{Moser:19, Schroff2023}, pupil aberrations\cite{Kaczorowski:16,Wojcik:26}, coherent stray light\cite{cerioni2026, Chen:24}, and partial source coherence\cite{Deng2017}. Conventional characterisation methods typically target a subset of these effects through dedicated metrology procedures such as interferometry, requiring additional hardware\cite{Zhou:22}, or characterising the components separately and reconfiguring the system\cite{DeMars:20}. However, subsequent system modifications introduce inconsistencies, as many of these properties are coupled in their effect on the field.

The emergence of differentiable optics has enabled \emph{in situ} characterisation by integrating the propagation models with gradient-based optimisation. Such methods are primarily implemented in 4F systems applied in near-field (Fresnel) holography for three-dimensional displays, where the main purpose of the CGH is to undo the weak diffraction at different reconstruction planes. In particular, camera-in-the-loop optimisation improved per-image experimental results\cite{Choi:21, Kim:25}, while neural-network-based hologram synthesis demonstrated generalisation to unseen targets, either under idealised propagation\cite{Shi2021} or with hardware non-idealities handled by calibrating the components separately\cite{Shi2022}. These approaches later converged into in-the-loop neural networks which achieved generalisation across unseen targets together with improved computational efficiency and image quality\cite{10.1145/3414685.3417802, 10.1145/3414685.3417846}.

Comparatively less attention has been paid to the far-field (Fraunhofer) regime, despite its importance in high-precision applications such as optical trapping\cite{PhysRevX.4.021034, app122010244} and lithography\cite{Bay:10, 4159977}. Unlike Fresnel systems, Fraunhofer propagation causes a full mixing of spatial modes, causing every degree of freedom in the modulation plane to influence the entire FoV\cite{Goodman2017}. The formed images therefore contain information about the complete optical system, posing an information-rich identification problem with long-range dependencies that are poorly suited to local operators used in Fresnel holography, such as convolutional neural networks.

Existing Fraunhofer approaches model only subsets of relevant physics and are limited to small Fourier-plane regions of interest (RoIs), resulting in degraded performance away from the optimisation region \cite{Schroff:24,10.1117/1.OE.64.9.094102}. Furthermore, these demonstrations have been shown to only address modest residual deviations in well-corrected optical systems that did not fundamentally alter the projected images. A complete system characterisation enabling accurate projection across the full \emph{first diffraction order}, corresponding to the entire uniquely controllable FoV of the Fourier plane, has not been demonstrated to date.

This work aims to close this gap. We investigate whether a compact, computationally efficient and physically interpretable digital twin can characterise a practical Fraunhofer holographic projector using hologram-image measurements alone and compensate for its dominant imperfections sufficiently to allow accurate control over the full bandwidth-limited field of view\cite{Matsushima:09}. To test the limits of this approach, the model is validated on a system exhibiting substantial deviations from nominal behaviour across multiple physical parameters. More broadly, we explore how far numerical propagation models can replace optical precision in practical holographic systems.

\section{Results}\label{sec:results}

\subsection{Modelling the optical system with a digital twin}

\begin{figure}[ht]
    \centering
    \includegraphics[width=\textwidth]{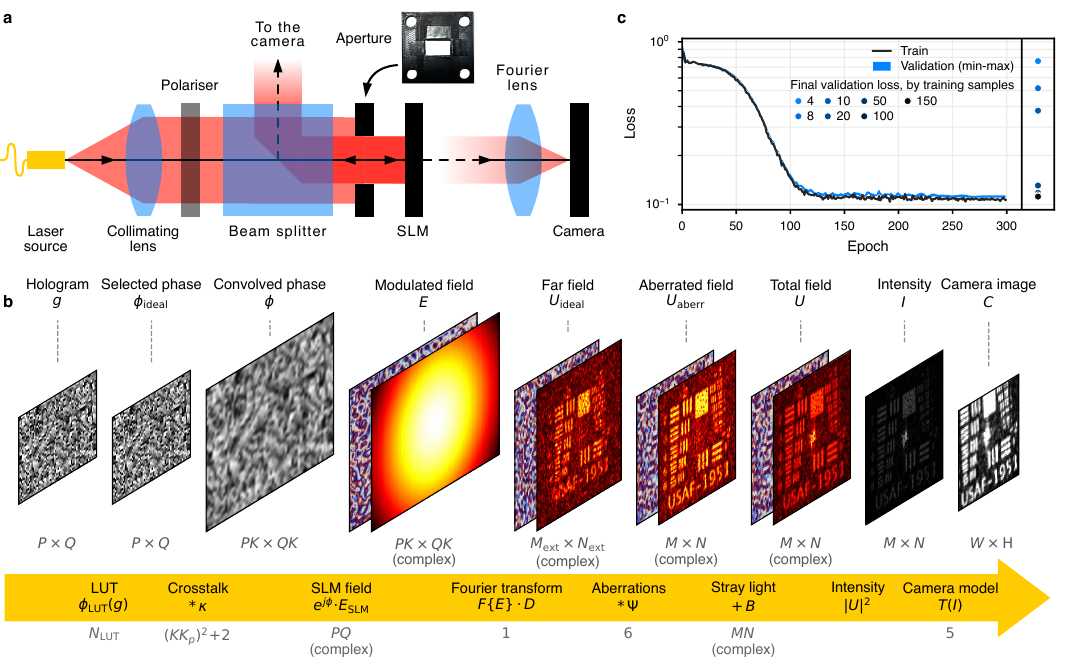}
    \caption{Overview of the optical system and the digital twin. (a) Diagram of the Fraunhofer holographic projection system, shown in an ``unfolded'' configuration along the optical axis. (b) Forward model of the digital twin. Intermediate stages are shown in sequence, with the corresponding operations, learnable parameters, and their dimensions annotated below each stage. (c) Convergence of the optical parameter identification stage for $N_S = 150$ samples, used for the remainder of the work. The column on the right compares final validation losses for alternative training-set sizes.}
    \label{fig:pipeline}
\end{figure}

The Fraunhofer holographic projector displays a greyscale \emph{hologram} $\mathbf{g}$ on a phase-only LCoS SLM and records the resulting focal-plane intensity pattern as an image $\mathbf{C}$ (Fig.~\ref{fig:pipeline}a). Ideally, the $(\mathbf{g}, \mathbf{C})$ are related by a Fourier transform, but in practice the field is modified by a sequence of physical effects along the optical path, each modelled as a differentiable operation with learnable parameters that together form a digital twin of the system (Fig.~\ref{fig:pipeline}b). The SLM converts $\mathbf{g}$ to phase modulation through an effective lookup table (LUT) $\phi_{\mathrm{LUT}}$, whose response is blurred between neighbouring pixels by fringing fields and liquid-crystal coupling modelled as a convolution with a \emph{crosstalk} kernel $\kappa$. The electrode fill factor $F$ and pixel deadspace reflectivity $\rho$ determine a brightness envelope $D$; the incident illumination $E_{\mathrm{SLM}}$ is complex-valued and generally non-uniform causing aberrations; and the Fourier lens imposes field-dependent aberrations through the pupil function $\Psi$. Coherent stray light $B$ interferes directly with the diffracted field, which is subsequently sampled by a camera whose coordinate system is related to the focal plane by an affine transform $\mathcal{T}$.

The twin is identified from a set of experimentally measured hologram-image pairs $(\mathbf{g},\mathbf{C})$ without additional metrology hardware or system modification (\nameref{sec:methods}). Training converges to a stable optimum (Fig.~\ref{fig:pipeline}c), overlapping with validation loss - computed continuously on a held-out set of patterns - indicating that the predictive power generalises.

In simulation, successful identification is achieved with as few as $N_S = 15$ hologram-image pairs and reaches the noise floor at approximately $50$ pairs (Supplementary Note~5). The training procedure is highly resilient to incomplete measurements, tolerating up to $6 \times$ Fourier-plane undersampling with full-FoV capture, first-order coverage as low as $16\%$ of the area when otherwise fully sampled, and more than $50 \times$ variation in exposure from severe underexposure to oversaturation. Experiments require between 100 and 150 pairs, likely due to unmodelled temporal variations such as brightness fluctuations, vibrations, air currents, and laser speckle, which limit the signal-to-noise ratio (SNR) of the data. 

\begin{figure}[ht]
    \centering
    \includegraphics[width=\textwidth]{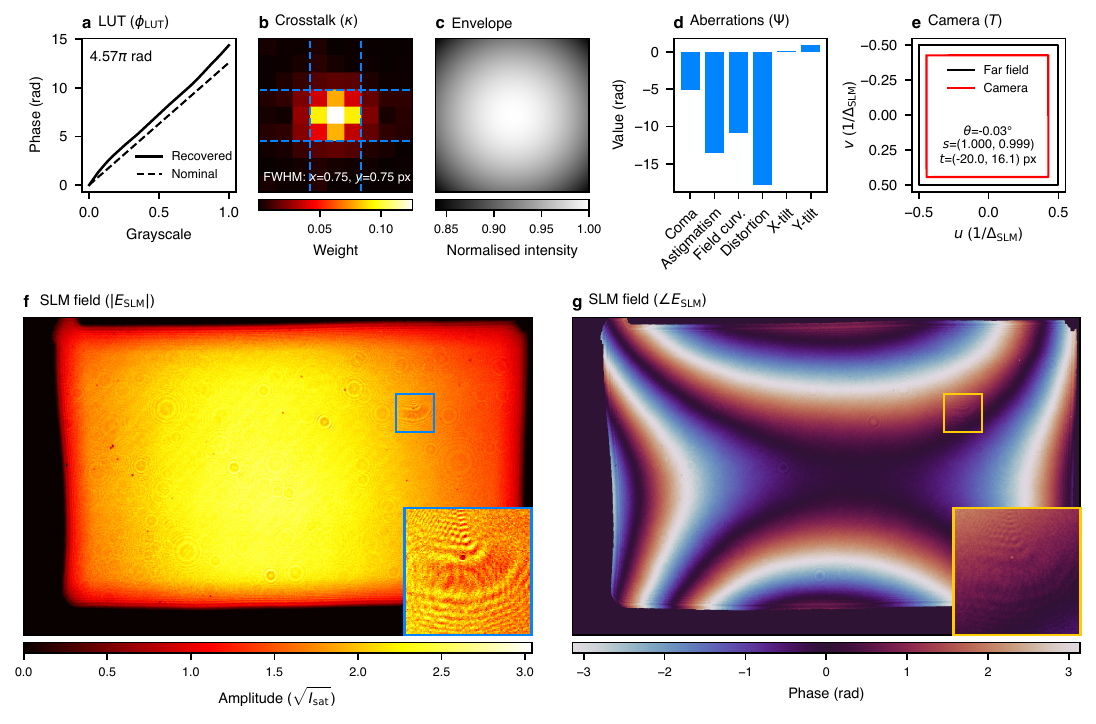}
    \caption{Identified optical parameters. (a) Phase LUT. (b) Crosstalk kernel, shown over the $K_p \times K_p = 3 \times 3$ pixel grid; blue lines mark the pixel boundaries. (c) Fourier-plane envelope. (d) Pupil aberration coefficients. (e) Affine transform: rotation $\theta$, scale $s$, translation $t$. (f, g) Amplitude and phase of the incident illumination field at each SLM pixel. The insets in (f) and (g) show diffraction fringes from dust in the system. The quantity $I_\text{sat}$ shown in (f) is the camera saturation intensity (Supplementary Note~1), which defines the amplitude scale.}
    \label{fig:recovered_params}
\end{figure}

The recovered parameters are shown in Fig.~\ref{fig:recovered_params}. The LUT deviates from the linear $4\pi$ response specified in the SLM's firmware, reaching a maximum phase modulation of $4.57\pi$ (Fig.~\ref{fig:recovered_params}a). The crosstalk kernel is nearly Gaussian, with a full width at half maximum of \SI{0.75}{\text{pixel}} along both axes, reaching neighbouring pixels - as we show below, this noticeably reduces the SLM's \emph{effective} modulation depth (Fig.~\ref{fig:recovered_params}b). The fill factor $F = 0.90$ was fixed to the datasheet value; the deadspace reflectivity was recovered as $\rho=0.91$ (Fig.~\ref{fig:recovered_params}c). Pupil aberrations are significant and spread across several Seidel terms - as expected for the wide incident beam passing through the doublet Fourier lens, but with negligible tip/tilt, indicating that the sensor is aligned with the Fourier plane\cite{Wojcik:26} (Fig.~\ref{fig:recovered_params}d). The recovered affine transform matches the expected system geometry, with a multiple-pixel shift and only a sub-pixel-scale rotation and near-unity scaling (Fig.~\ref{fig:recovered_params}e). The recovered illumination amplitude and phase (Fig.~\ref{fig:recovered_params}f,g) represent deviations from an ideal plane wave. They absorb the cumulative effect of incident wavefront, aperture clipping, dust contamination, and their diffraction effects\cite{Wojcik2025}. A dark region with diffraction fringes along the bottom and left edges is caused by a misplaced aperture intended to suppress stray reflections. This shows that the true illuminated extent of the SLM can be identified. At full pixel resolution, the field also resolves diffraction ringing from individual dust particles, with the insets showing both point-like and elongated features consistent with contamination on and above the SLM surface. The recovered stray light is discussed separately as its additive nature complicates joint optimisation with the other parameters.

\subsection{Effects of identified parameters on image quality}

\begin{figure}[ht]
    \centering
    \includegraphics[width=\textwidth]{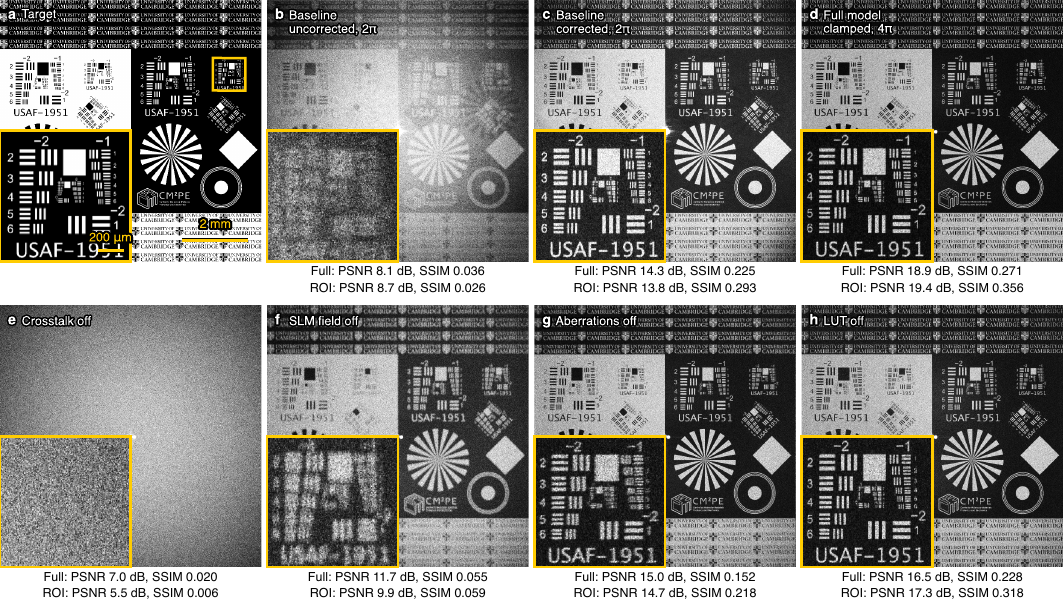}
    \caption{Effects of physical parameters on the formed image, captured experimentally. All images are time averages of $N_B = 4$ jointly optimised holograms and are presented on a linear brightness scale from no light (black) to camera saturation (white). The insets show the same Fourier-plane region for all cases, as marked in panel (a). PSNR and SSIM metrics are reported for both the full FoV and the inset region. (a) Target pattern. (b, c) With $2\pi$ phase modulation and (b) idealised and (c) identified system parameters. (d) With $4\pi$ phase modulation and the identified system parameters. (e)-(h) Ablation study of the identified system parameters on the $4\pi$ phase modulation, where each parameter group is individually replaced with its idealised value: (e) phase crosstalk kernel, (f) illumination field, (g) pupil function, (h) LUT.}
    \label{fig:ablation}
\end{figure}

The identified models are used to synthesise holograms which recreate target intensity distributions that span the entire first diffraction order and to predict the corresponding camera images (\nameref{sec:methods}). The SLM firmware allows for adjustment of the nominal phase modulation range. At the standard $2\pi$ setting, CGH under idealised system parameters produces a heavily distorted image relative to the target (Fig.~\ref{fig:ablation}a,b): non-uniform illumination and pupil aberrations blur and distort the image, while the large crosstalk creates a superimposed ghost copy rotated \SI{180}{\degree} from the intended pattern and increases speckle-like noise. The $2\pi$ phase modulation configuration was characterised independently (Supplementary Note~4), producing sharper reconstructions with improved metrics and no ghost replica (Fig.~\ref{fig:ablation}c). However, severe crosstalk reduces the effective phase modulation range, resulting in a non-uniform brightness envelope and reduced usable FoV. This effect is most pronounced at the corners of the Fourier plane, where the highest spatial frequencies require the steepest phase gradients on the SLM. The CGH process can mitigate the resulting noise and ghosting, but not this fundamental limitation. To overcome it, the SLM was operated at nominal $4\pi$ phase modulation depth with the resulting digital twin, discussed in the previous section, used throughout the remainder of this work, producing substantially more uniform reconstruction across the FoV (Fig.~\ref{fig:ablation}d) and improved metrics.

The modular structure of the trained twin is used in an ablation study by substituting one identified parameter at a time with its ideal value during hologram synthesis, with all others fixed at their identified values. Starting from the most severe defect, replacing the recovered crosstalk kernel $\kappa$ with an ideal delta function turns the projection into noise (Fig.~\ref{fig:ablation}e). At modulation depth of $\SI[parse-numbers=false]{4\pi}{\radian}$, uncompensated crosstalk strongly couples neighbouring pixels, scrambling their phases around the complex unit circle which destroys the intended profile. The next dominant effect arises from illumination $E_{\mathrm{SLM}}$, as its complex-valued nature mixes the desired phase and amplitude through convolution with the resulting point-spread function (PSF), introducing both blur and additional noise. This is followed by pupil aberrations $\Psi$ (Fig.~\ref{fig:ablation}g), which produce similar effects in a field-dependent manner. Finally, despite its $14\%$ deviation from its nominal $4\pi$ phase range, $\phi_{\mathrm{LUT}}$ (Fig.~\ref{fig:ablation}h) has the smallest effect, increasing noise and reducing contrast. Neglecting the envelope $D$ (not shown) causes a smooth intensity variation across the image (Fig.~\ref{fig:recovered_params}c).

\subsection{Full first-diffraction-order projection}

\begin{figure}[ht]
    \centering
    \includegraphics[width=\textwidth]{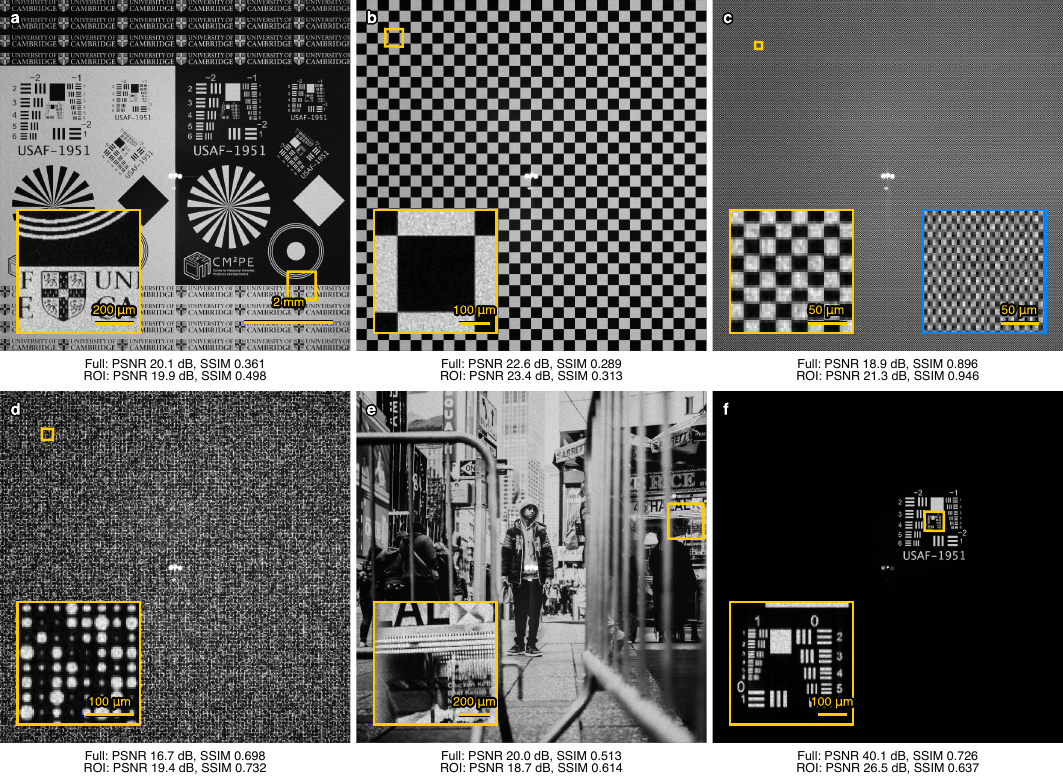}
    \caption{Captured reconstructions of full first-order FoV target patterns, with selected details shown in the insets. All images are time averages of $N_B = 20$ holograms. PSNR and SSIM metrics for the full FoV captures and the insets are reported under each panel. (a) Complex binary pattern. (b) Coarse chequerboard. (c) Fine chequerboard. The blue inset on the right shows the same region for a denser chequerboard ($1280 \times 800$ tiles) to show that bandwidth limit can be approached. However, the camera undersamples the field, so this is not analysed further. (d) Aperiodic Huygens metasurface-like pattern. (e) Greyscale image, sourced from Pexels (photo ID 20988017). (f) Small USAF-1951 resolution chart, adapted from Wikimedia Commons (USAF-1951.svg, Setreset, CC BY-SA 4.0). }
    \label{fig:target_study}
\end{figure}

Figure~\ref{fig:target_study} shows multiple captured patterns spanning the full first diffraction order of the SLM, \SI{7.91}{\milli\metre} $\times$ \SI{7.91}{\milli\metre} in area. The chosen image scale balances full-FoV coverage against resolving the smallest features supported by the camera's \SI{3.45}{\micro\meter} pixel pitch.

Projecting binary patterns (Fig.~\ref{fig:target_study}a-d) is challenging but especially revealing, as uniform intensity must be maintained over the entire FoV. The inset in Fig.~\ref{fig:target_study}a shows pixel-level features in the circular pattern, demonstrating that the system approaches the spatial-bandwidth limit of the SLM. The square chequerboard images (Fig.~\ref{fig:target_study}b,c) show uniformity at two spatial frequencies, with 120-pixel and 10-pixel pitches - the latter giving $384 \times 384$ individual tiles in a single projection, relevant for applications such as optical-tweezer arrays. At finer tested pitches, the camera pixel size begins to undersample the intensity, causing blur relative to the target. Fine, aperiodic structures are also possible: the image in Fig.~\ref{fig:target_study}d is a Huygens metasurface-like pattern with circular features of varying size on a $256 \times 256$ cell grid, signifying the potential relevance of this approach to simultaneous large-area exposure in holographic lithography. The projection is not limited to binary content: high-quality greyscale images can also be formed (Fig.~\ref{fig:target_study}e). The method also supports localised targets, similar to those used in typical RoI-limited studies: Fig.~\ref{fig:target_study}f shows a small USAF resolution target. Because the total optical power is fixed by the laser source, concentrating it into a smaller area increases the achievable SNR and contrast. Consequently, the smaller USAF target was captured at \SI{10}{\milli\second} exposure, while full-FoV images required \SI{400}{\milli\second} for comparable brightness, which makes such projections naturally more sensitive to residual noise and system vibrations.

\begin{figure}[ht]
    \centering
    \includegraphics[width=\textwidth]{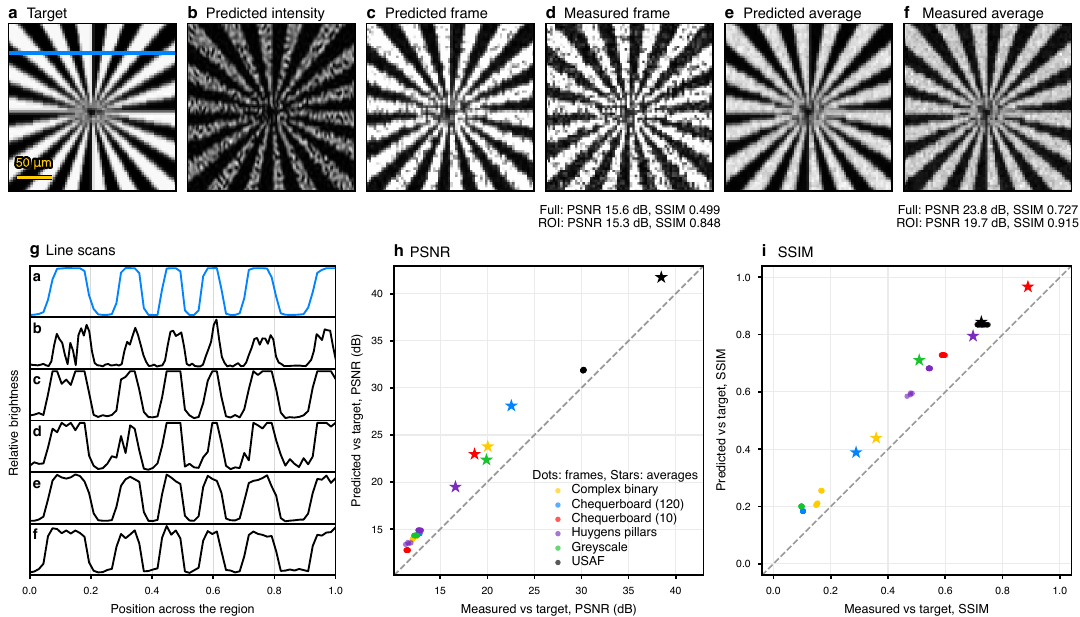}
    \caption{Effects of the camera on the observed image, shown for the same selected region of the complex binary pattern (Fig.~\ref{fig:target_study}a). Predicted images are obtained by passing the displayed holograms through the identified digital twin and its camera model. (a) Target intensity passed through the camera model. The blue line indicates the line scans shown in (g). (b)-(d) Single-frame results: (b) simulated intensity, (c) simulated camera image, and (d) experimental image. (e, f) Time-averaged camera images: (e) simulated and (f) experimental. Panels (d) and (f) report metrics between the simulated and experimental image pairs. (g) Corresponding line scans, normalised to show relative variation. (h, i) Comparison of the target-measured ($x$-axis) and the target-predicted ($y$-axis) PSNR and SSIM metrics, respectively, to show correlation between experiments and predictions.}
    \label{fig:camera_model}
\end{figure}

Quantitative evaluation is fundamentally limited because the Fourier-plane intensity is filtered by the camera response (Fig.~\ref{fig:camera_model}). This is particularly apparent when projecting individual holograms: the intensity can span a large dynamic range (Fig.~\ref{fig:camera_model}b), which becomes clipped by the camera response (Fig.~\ref{fig:camera_model}c,d). The predicted and experimental images exhibit closely matching noise patterns, indicating that the CGH process is responsible for a significant portion of the noise. The remaining discrepancies, most visible in the dark regions, represent a genuine residual twin-experiment mismatch. Residual brightness differences arise primarily from fluctuations in laser polarisation, which change the transmitted optical power and the degree of saturation clipping. Time averaging over $N_B = 20$ holograms (Fig.~\ref{fig:camera_model}e,f) suppresses stochastic CGH noise and substantially improves agreement between prediction and experiment. This indicates that the twin captures the dominant systematic errors in the optical system, while the remaining discrepancies are mainly due to CGH noise, also supported by the line scans in Figure~\ref{fig:camera_model}g. However, the digital twin systematically overestimates image quality, predicting higher PSNR and SSIM\cite{1284395} values than those observed experimentally (Fig.~\ref{fig:camera_model}h,i). This trend remains consistent across targets and averaging levels, indicating stable relative performance in predictive power. The differing order of the metric rankings underscores their limitations and the need for caution in interpreting absolute values, reflecting a known challenge in fair holographic quality assessment \cite{Yang2022}.

\subsection{Measurement and suppression of coherent stray light}

\begin{figure}[ht]
    \centering
    \includegraphics[width=\textwidth]{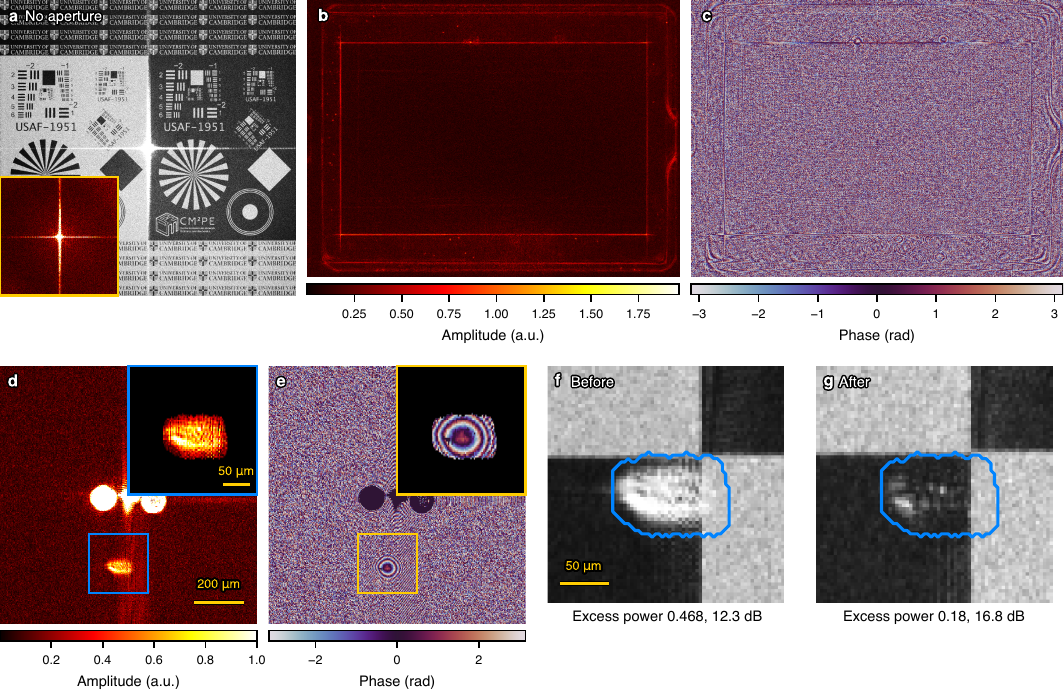}
    \caption{Study of the stray light $B$ and its effects on the reconstructed image. (a) Complex image projected without an aperture in front of the SLM. The inset shows recovered amplitude of $B$. (b, c) Amplitude and phase of the Fourier transform of the recovered $B$ with the aperture removed from the system. (d, e) Central region of the remaining $B$ after the aperture is placed in front of the SLM. The insets show the RoI used in the stray-light correction procedure. (f, g) Observed image in the aperture case, before and after stray-light correction. The blue outline marks the region used in the correction procedure.}
    \label{fig:stray_light}
\end{figure}

The reconstructions presented above do not compensate for the stray-light field $B$, which is studied separately in Figure~\ref{fig:stray_light}. To determine its physical origin, a separate experiment was performed with the aperture - otherwise present throughout this work - removed from the system. Without the aperture, $B$ is dominated by a bright central cross that spans the entire FoV (Fig.~\ref{fig:stray_light}a). Its Fourier transform (Fig.~\ref{fig:stray_light}b,c) reveals that $B$ is caused by reflections from the SLM's inactive surface surrounding the rectangular addressable area that controls modulation. $B$ can only be determined where the probe images are not saturated (the edges of the cross), so this Fourier transform is only partially recovered, capturing aperture edge effects instead of the full smooth profile. Stray light is often attributed to the SLM fill factor, but the recovered field in Fig.~\ref{fig:stray_light}b,c shows a clear structure with diffraction fringes from individual dust particles, indicating that it is dominated by reflections from the inactive SLM surface surrounding the pixel area, a mechanism not captured by the fill factor model.

When installed, the aperture removes the cross-shaped artefact, leaving only a small residual $B$ (Fig.~\ref{fig:stray_light}d,e). The two visible circular structures come from spurious reflections within the beam splitter, saturating the camera and preventing phase recovery. The relatively low SNR in the reported images  happens because $B$ additively absorbs residual mismatch between the predicted and observed training data. For a proof-of-principle correction, a region with resolved structure was selected (insets in Fig.~\ref{fig:stray_light}d,e), with the recovered $B$ incorporated directly into the CGH process. The resulting images (Fig.~\ref{fig:stray_light}f,g) show more than a $2.5 \times$ reduction in the power of the localised stray light, with the remaining residual likely limited by the SNR.

\section{Discussion}\label{sec:discussion}

The recovered digital twin enables bandwidth-limited image projection over the entire first diffraction order. To our knowledge, this is the first report of such holographic reconstructions in the Fraunhofer regime, with complex, high-resolution patterns presented as complete images rather than cropped regions of interest or simulation alone. Repeated identification under different aperture and modulation-depth configurations (Supplementary Note~4) demonstrates the robustness of the training procedure. Once identified, the twin is used without further measurements or refinement, requiring recalibration only if the optical system is modified or is expected to drift with time.

Ultimately, the strongest validation for the digital twin is not the particular values of its parameters or any single image-quality metric, but its ability to generalise beyond the training data by experimental synthesis of previously unseen holograms. The full-FoV performance is enabled by explicit distinction between the illumination field $E_\text{SLM}$ and pupil aberrations $\Psi$. Because $E_\text{SLM}$ multiplies the field in the SLM plane (Eq.~(\ref{eqn:diff_model})), the resultant PSF convolves the ideal projection causing uniform global blur and noise. By contrast, $\Psi$ causes this PSF to vary within the field, introducing field-dependent distortions that cannot be represented by $E_\text{SLM}$ alone. Previous Fraunhofer identification approaches neglected $\Psi$ and as a result achieved only limited-RoI performance \cite{Schroff:24,10.1117/1.OE.64.9.094102}. This occurs because $E_\text{SLM}$ absorbs aberrations when matching small RoIs which fails to generalise across the full FoV. Only joint identification of $E_\text{SLM}$ and $\Psi$ captures field dependence and enables accurate full-field projection (Supplementary Note~5). 

The recovered stray light is dominated by reflection from the inactive surface surrounding the addressable area, rather than the commonly assumed pixel-deadspace source. This implies that a model parametrising only the fill factor is structurally unable to represent such dominant stray-light contribution, regardless of how well its parameters are fitted. To the best of our knowledge, this is the first direct identification of this parameter within a holographic calibration framework, contributing to the understanding and compensation of residual zeroth-order light. Previous work \cite{cerioni2026} demonstrated zeroth-order suppression only for the central bright spot corresponding to the zeroth order, and only within sparse patterns consisting of a few isolated features. In contrast, the correction presented here operates within rich images spanning the full FoV, although it is currently limited to selected high-SNR regions. Improved optimisation methods and higher-SNR measurements should extend it across larger areas of the Fourier plane, further reducing the need for precise alignment and dedicated stray-light-suppression optics.

Some parameters are not uniquely identifiable. Without an absolute reference, linear phase ramps in $E_\text{SLM}$ and affine shifts in $\mathcal{T}$ produce identical translations of the observed image, which may shift the images relative to their intended locations. Similarly, the decomposition of the envelope $D$ into $F$ and $\rho$ is not unique, so $F$ is fixed to maintain identifiability (Supplementary Note~2). Therefore, the value of $\rho$ should be interpreted as a description of $D$ instead of a direct reflectivity measurement.

The model has two main limitations, both of which stem from the discrete implementation rather than the underlying physics. First, for sufficiently severe aberrations, the local PSFs move beyond or exceed the isoplanatic tiles used in computing $\Psi$ which reduces the approximation accuracy. Second, time averaging is required to achieve high image quality by averaging noise caused by optical vortices (dark spots in Fig.\ref{fig:camera_model}(b)). Existing vortex-suppression techniques either rely on invertible propagation models\cite{Schroff2023}, which are incompatible with the present approach, or are limited to small Fraunhofer RoIs ($\approx 10\%$ FoV)\cite{10.1117/1.OE.64.9.094102} despite being effective in the Fresnel domain\cite{Zhu2024}. Extending vortex-free CGH to full-FoV projection while remaining robust to crosstalk and significant field-dependent aberrations remains an important direction for future work.

The ability to control the full FoV is particularly valuable for applications involving large numbers of independently addressable points over extended areas. In the demonstrated system, approximately $3840 \times 2400$ Nyquist-limited Fourier-plane samples can be controlled simultaneously. Due to the Fourier-transform relationship in the Fraunhofer regime, this capability is largely scale-invariant down to the diffraction limit, making it directly applicable to large optical-trap arrays and holographic lithography. Beyond hologram synthesis, the model provides a framework for evaluating CGH algorithms in practical optical hardware rather than idealised propagation environments. Future work should further reduce the remaining discrepancies while preserving the computational efficiency and physical interpretability of the approach.

\section{Methods}\label{sec:methods}

\subsection{Optical system and propagation model}

The holographic projector is composed of a $\lambda = \SI{632.8}{\nano\meter}$ HeNe laser source (Melles Griot 05-LHP-151), a phase-only LCoS SLM (Santec SLM-200, $P \times Q = 1920 \times 1200$ pixels of \SI{8}{\micro\meter} pitch), Fourier lens (Thorlabs AC254-100-A, $f=\SI{100}{\milli\meter}$), and a monochrome CMOS camera (Kiralux CS126MU, \SI{3.45}{\micro\meter} pixel pitch) (Fig.~\ref{fig:pipeline}a). The beam was collimated with a second AC254-100-A lens, passed through a polarising beamsplitter cube (Thorlabs CCM1-PBS251/M) to align the selected polarisation with SLM's axis, and a non-polarising beamsplitter (Thorlabs CM-BS013) separated the incident and reflected optical paths. The SLM was operated in $2\pi$ and $4\pi$ phase-modulation modes, depending on the experiment and specified in its firmware. The projected first diffraction order forms a square of side length \SI{7.91}{\milli\meter}, with band-limited $M \times N = 3840 \times 2400$ sampling of size \SI{2.06}{\micro\meter} $\times$ \SI{3.30}{\micro\meter} along the horizontal and vertical axes, respectively. The central $W \times H = 2000 \times 2000$ pixel camera region is used for training, increasing to $3000 \times 3000$ pixels for CGH measurements, to show that the identified model remains effective beyond the training window.

The proposed model consists of two parts: $\mathcal{M}(\mathbf{g}) = I$ predicts the Fourier plane intensity $I$ given the greyscale hologram $\mathbf{g} \in [0,1]^{P \times Q}$; $\mathcal{C}(I) = \mathbf{C}$ predicts the greyscale camera image in the normalised range $\mathbf{C} \in [0, 1]^{W \times H}$, in units of camera saturation corresponding to intensity $I_\text{sat}$. The model is described by a sequence of differentiable relations (Fig.~\ref{fig:pipeline}b), with the complete derivation and treatment presented in Supplementary Note~1. 

The phase modulation in the SLM plane $(x,y)$ is computed with a monotonic function $\phi_{\mathrm{LUT}}$ and convolution with an inter-pixel crosstalk kernel $\kappa$, resulting in the phase profile
\begin{equation}
    \phi(x,y) = \phi_{\mathrm{LUT}}(\mathbf{g}) * \kappa.
\end{equation}
$\phi_{\mathrm{LUT}}$ is constructed by $N_\text{LUT} = 20$ equally spaced nodes which define non-negative local gradients. $\kappa$ consists of a centred Gaussian component, parametrised by its standard deviations along both axes, and a free residual sampled over $K_p \times K_p$ pixel support, each with a $K \times K$ sub-pixel grid, with $K = K_p = 3$ proving sufficient in the experiments. The convolution is performed on this sub-pixel grid by first upsampling the other fields. The phase modulates the incident field on the SLM, $E_{\mathrm{SLM}}(x,y)$, defined on the $P \times Q$ pixel grid. The field is additionally modulated by the aperture function of the SLM pixel grid, $A_{\mathrm{SLM}}(x,y)$, described by the electrode fill factor $F$ and the deadspace reflectivity $\rho$. Its effect is included in the model using the resultant multiplicative brightness envelope $D$ in the Fourier-plane $(u,v)$. The reflected field is propagated to the Fourier plane using a diffraction model that includes field-dependent aberrations in the pupil function $\Psi(x,y;u,v)$ parametrised by $6$ coefficients (coma, astigmatism, field curvature, distortion, and $x$ and $y$-tilts) modelled by a bilinear interpolation over $N_\text{tile} \times N_\text{tile} = 40 \times 40$ isoplanatic tiles, as in Ref.~\citenum{Wojcik:26}. The field interferes with coherent stray light $B$ defined in the Fourier plane, sampled on the $M \times N$ grid. The full propagation model is described as
\begin{equation} \label{eqn:diff_model}
U(u,v) = B(u,v) + \iint \Psi(x,y;u,v)~E_{\mathrm{SLM}}(x,y) ~A_{\mathrm{SLM}}(x,y)~e^{j\phi(x,y)} e^{-2\pi j(ux+vy)} \,\mathrm{d}x\,\mathrm{d}y.
\end{equation}
The camera samples the intensity $I = |U|^2$ and is modelled by affine misalignment $\mathcal{T}$ between the optical and sensor coordinate systems, integration over the pixel area and exposure time, and saturation to compute $\mathbf{C}$. $\mathcal{T}$ is described by a rotation $\theta$ relative to the $u$-axis, shift $\mathbf{t}$ from the optical axis, and a scale $\mathbf{s}$ along both axes relative to the expected camera coverage within the first diffraction order. The bottom of Figure~\ref{fig:pipeline}b summarises the trainable parameters and their numerical sizes. Additional implementation details are provided in Supplementary Note~2.

\subsection{Training procedure}
The digital twin is trained from experimental hologram-capture pairs $(\mathbf{g},\mathbf{C}_\text{exp})$ by matching $\mathbf{C}_\text{exp}$ to the predicted images $\mathbf{C}_\text{pred} = \mathcal{C}(\mathcal{M}(\mathbf{g}))$ (Supplementary Note~3). In terms of $\mathbf{g}$, the dataset consists of $4$ binary chequerboard gratings used for the initial estimation of $\mathcal{T}$; $1$ binary chequerboard with minimal (2 pixel) period - it deflects light into the corners of the first order, revealing the zeroth order in the centre - it is thresholded so that $\mathbf{C}_\text{exp} < 0.1$ serves as the mask $\mathbf{M}_\text{ZOD}$ below; $N_S = 150$ random greyscale holograms used for optical parameter identification; $10$ more random holograms for convergence validation (Fig.~\ref{fig:pipeline}c). The captures of gratings used \SI{50}{\milli\second} exposure time and the random noise probes \SI{200}{\milli\second}.

Following data acquisition, no additional measurements are used during training or hologram synthesis. Training proceeds in four stages. Stages~3 and 4 use the AdamW\cite{loshchilov2019decoupledweightdecayregularization} optimiser to minimise loss, 
\begin{equation}
\mathcal{L}_\text{train}(\mathbf{C}_\text{pred},\mathbf{C}_\text{exp};\mathbf{M}) =
\frac{1}{2} \left[1-\mathrm{SSIM}(\mathbf{C}_\text{pred},\mathbf{C}_\text{exp})\right]_{\mathbf{M}}
+ \frac{1}{2} \left.\frac{\langle (\mathbf{C}_\text{pred}-\mathbf{C}_\text{exp})^2\rangle}{\langle \mathbf{C}_\text{exp}^2\rangle}\right|_{\mathbf{M}},
\end{equation}
evaluated in a training batch of $N_T$ image pairs, where $\langle\cdot\rangle$ denotes the mean over the pixels retained by mask $\mathbf{M}$ and over the batch, and SSIM is also averaged over the batch.

\begin{enumerate}
\item \textbf{Initialisation.}
All parameters are initialised with nominal values with small random perturbations to avoid training degeneracies (Supplementary Note~3).

\item \textbf{Affine estimation.}
The affine transform is estimated first to remove the geometric ambiguity before joint optimisation. The chequerboard-based measurements project uniform grids of points in the Fourier plane. Camera saturation is neglected in the simulated images at this point, to decouple its effect from the geometry estimation. The measured and simulated point grids are matched in normalised Fourier-plane coordinates, and $\mathcal{T}$ is fitted using RANSAC\cite{10.1145/358669.358692}.

\item \textbf{Optical parameter identification.} (300 epochs)
All parameters other than $B$ are jointly optimised, with $B=0$, to minimise $\mathcal{L}_\text{train}(\mathbf{C}_\text{pred},\mathbf{C}_\text{exp};\mathbf{M}_\text{ZOD}) + \mathcal{R}(E_\text{SLM})$. $\mathbf{M}_\text{ZOD}$ excludes the undiffracted zero-order region dominated by $B$, and $\mathcal{R}$ computes total-variation and Laplacian regularisation on the amplitude and phase of $E_\text{SLM}$ (Supplementary Note~3). Afterwards, the amplitude of $E_\mathrm{SLM}$ is thresholded to isolate the largest connected illuminated region, and the field is smoothly masked to zero outside it, removing residual low-amplitude noise otherwise mistaken for signal.

\item \textbf{Stray light estimation.} (100 epochs)
All parameters except $B$ are fixed and optimised to minimise $\mathcal{L}_\text{train}(\mathbf{C}_\text{pred},\mathbf{C}_\text{exp};\mathbf{M}_\text{sat}) + \mathcal{R}(\mathcal{F}^{-1}(B))$, where $\mathbf{M}_\text{sat}$ excludes pixels that appear saturated in every image, and $\mathcal{F}^{-1}$ is the inverse Fourier transform used to constrain the SLM-plane structures in $B$. Afterwards, $B$ is set to unit amplitude and zero phase where its amplitude exceeds the saturation level, as such pixels would otherwise bias the estimate despite containing no additional information.
\end{enumerate}

The implementation is designed to work with limited computational resources. Training used $N_S = 150$ $(\mathbf{g},\mathbf{C})$ pairs, batch size $N_T = 4$ in micro-batches of $2$ pairs, on a single NVIDIA GTX 1080 GPU (\SI{8}{\giga\byte} VRAM), with each stage using a constant learning rate for the first $50\%$ of epochs, followed by cosine decay to zero (Supplementary Note~3). Stage~3 converged within the first $150$ of its $300$ epochs (\SI{24}{\minute} total); stage~4 took \SI{5}{\minute}.

The decision to use random probes in stages~3 and~4 was to avoid bias towards specific parameter values while efficiently exciting the full spatial bandwidth of the optical system. However, the training landscape is highly non-convex and the resulting images are largely isotropic, so direct optimisation with unknown $\mathcal{T}$ tends to converge to shallow local minima, with optical parameters fitting unphysical noise configurations instead of the true image registration. Stage~2 therefore uses periodic binary chequerboards, whose Fourier spectra form well-separated spots ideal for registration grids and remain robust to severe system defects, including strong crosstalk, due to the large uniform phase regions they contain.

Similarly, the coherent field $B$ is estimated only after the remaining optical parameters have converged. Because $B$ enters $U_{\mathrm{aberr}}$ additively, joint optimisation could allow it to absorb unrelated modelling errors and bias the remaining parameters. Once the other parameters have settled, the interference between the static field $B$ and the random probe patterns provides sufficient information to recover both its amplitude and phase.

Stage~3 uses multi-scale SSIM\cite{1292216} loss in the implementation of $\mathcal{L}_\text{train}$ as we found that single-scale SSIM, used in stage~4, stagnates pupil estimation - likely because lens aberrations vary over large scales. The Seidel distortion term in $\Psi$ is correlated with the affine scaling, but $\mathcal{T}$ applies uniform scaling and $\Psi$ is radially nonlinear, so joint refinement in stage~3 resolves this ambiguity. The standard pyramid weights of multi-scale SSIM, which are set up for human perception, were modified to increase significance of higher spatial frequencies while retaining non-zero coarse-scale contributions (Supplementary Note~2). 

\subsection{Hologram generation}
Time-Multiplexed Batch Gradient Descent\cite{10.1117/12.3040563} is used with the trained optics model $\mathcal{M}$ to jointly optimise batches of holograms $\mathbf{g}_i$. Depending on the experiment, $N_B = 4$ or $20$ holograms are used, where for $N_B > 4$, the holograms are optimised independently in batches of $4$ and concatenated for sequential projection. The batches minimise the normalised mean squared error (NMSE) between their time-averaged intensity $\bar{I}$ and the target distribution $T$,
\begin{equation} \label{eqn:NMSE_avg}
\mathcal{L}_\text{NMSE}(I,T) = \frac{\langle (\tilde{I}-\tilde{T})^2\rangle}{\langle \tilde{T}^2\rangle},
\end{equation}
where $\tilde{X} = X/\langle X\rangle$ and $\langle\cdot\rangle$ is the spatial mean, and
\begin{equation} \label{eqn:int_avg}
\bar{I} = \frac{1}{N_B}\sum_{i=1}^{N_B}\mathcal{M}(\mathbf{g}_i).
\end{equation}
This normalisation removes sensitivity to absolute brightness, allowing the optimisation to focus on matching the overall distribution. The camera stage $\mathcal{C}$ is neglected during CGH optimisation and all trained parameters are fixed. $T$ is prepared by loading a square image with side length $\max(M,N)$ which spans the entire first diffraction order, and bilinear interpolation to the Nyquist-limited resolution $M\times N = 2P \times 2Q$\cite{Matsushima:09}. The batch of $\mathbf{g}_i$ (each hologram of $P \times Q$ pixels) is generated jointly using AdamW optimisation in $300$ epochs at learning rate $5 \times 10^{-2}$ and weight decay $10^{-4}$, using cosine learning rate decay to $0$.

During optimisation, $\mathbf{g}_i$ is clipped to $[0,1]$ after each update to enforce valid SLM values. After optimisation, $\mathbf{g}_i$ is quantised to $1024$ integer greyscale levels required by the 10-bit SLM driver. For the idealised propagation model assuming $2\pi$ phase modulation, the phase $\phi$ was instead optimised directly and wrapped to the interval $[0,2\pi)$ after each update using a modulo-$2\pi$ operation, which empirically improved optimisation performance. Such wrapping cannot be assumed for the identified system, where the range of $\phi_\text{LUT}$ is not an integer multiple of $2\pi$, so clipping is used there instead. 

During stray-light suppression, a mask is manually selected over a region of relatively high SNR from the recovered $B$, setting $B$ to zero elsewhere. The remaining $B$ is included in the forward model and used during hologram synthesis. The reported ``excess power'' is computed as the mean squared error between the measured image within the captured region and the scaled target image, as described below.

The presented images are created by frame averaging in post-processing, where images $\mathbf{C}_i$ are captured for each $\mathbf{g}_i$ separately and averaged to $\bar{\mathbf{C}}$ in the same fashion as in Eq.~(\ref{eqn:int_avg}) to avoid an undetermined SLM state during switching\cite{Goetz_2013}, which was found to be particularly severe for the tested SLM. In practice, this can be implemented with a single camera exposure by extending the exposure time relative to the hologram switching time, or by gating with a shutter during switching. 
 
Reported metrics in image comparisons are computed in the camera pixel grid, over the \emph{valid} region defined as all of the first-order area excluding a circle of $40$-sample radius in the $M \times N$ target space, the location of the zeroth order that would otherwise bias all the metrics. The images $\bar{\mathbf{C}}$ are normalised so that the dynamic range of the camera (saturation) is $1$. For each, the target image is affine-transformed to align with $\bar{\mathbf{C}}$ and resampled using the camera model to $\mathbf{C}_T = \mathcal{C}(T)$, after which it is scaled to $\hat{\mathbf{C}}_T = \alpha \mathbf{C}_T$ to match $\bar{\mathbf{C}}$ with an optimal least-squares scalar fit $\alpha$. The valid region is also transformed to the camera pixel grid. The peak signal-to-noise ratio (PSNR) is
\begin{equation}
    \mathrm{PSNR} = 10\log_{10}\!\left(\frac{\text{data range}}{\langle (\bar{\mathbf{C}} - \hat{\mathbf{C}}_T)^2\rangle}\right),
\end{equation}
where ``data range'' is the maximum - minimum value of the images $\bar{\mathbf{C}}$ and $\hat{\mathbf{C}}_T$ within the valid region, with $\langle\cdot\rangle$ computed only over the valid region. The reported SSIM metrics are evaluated on the same image pairs using the same valid-region mask.

\subsection{Disclosure}
Anthropic Claude models were used to suggest improvements to manuscript structure, to assist in preparing the figures, and to refactor the code for release. The presented methods, experiments, analysis, and presentation decisions were made by the authors, who wrote the text, reviewed the figures and code, and take full responsibility for their correctness.

\subsection{Code availability}
The code used to generate the results in this work is available on GitHub, \url{https://github.com/antoni-wojcik/holo-twin}.

\subsection{Data availability}
The datasets generated and analysed in this study are available  in the Apollo -- University of Cambridge repository, \url{https://doi.org/10.17863/CAM.134536}.

\subsection{Supplementary information}
Supplementary Information is appended to this manuscript, following the main text.

\section{Acknowledgements}
This work was supported by the EPSRC Centre for Doctoral Training in Connected Electronic and Photonic Systems [EP/S022139/1] and by EPSRC [EP/V055003/1].

\section{Author contributions}
A.J.W. conceived the idea and study, developed the methodology, performed all experiments, analysed the data, and wrote the manuscript. D.S. contributed to the development of the methodology, the experimental setup, and the interpretation of the data. T.D.W. and H.J.J. provided the materials and supervised the research. All authors reviewed the manuscript.

\section{Conflict of Interest}
The authors declare no conflict of interest.

\clearpage
\setcounter{page}{1}     \renewcommand{\thepage}{S\arabic{page}}
\setcounter{figure}{0}   \renewcommand{\thefigure}{S\arabic{figure}}
\setcounter{table}{0}    \renewcommand{\thetable}{S\arabic{table}}
\setcounter{equation}{0} \renewcommand{\theequation}{S\arabic{equation}}
\setcounter{section}{0}
\setcounter{footnote}{0}
\def\citenumfont#1{S#1}
\def\bibnumfmt#1{[S#1]}

\begin{center}
  {\LARGE Supplementary Information for ``Digital twin for full field-of-view Fraunhofer holographic patterning''\par}
  \vspace{1em}
  {\large Antoni J. Wojcik, Dilawer Singh, Hannah J. Joyce, and Timothy D. Wilkinson\par}
\end{center}
\vspace{1em}

\section{Supplementary Note 1: Forward model} \label{sec:forward}
\subsection*{Derivation of the continuous equations}
This section derives the forward model used in the main manuscript, relating the greyscale values applied to the SLM to the image recorded by the camera. A complete electromagnetic model of light propagation, including diffraction, polarisation, and liquid-crystal (LC) dynamics, is computationally prohibitive for optimisation. Instead, we develop a simplified model that captures the dominant physical effects and remains computationally efficient.

\begin{figure}[ht]
    \centering
    \includegraphics[width=\textwidth]{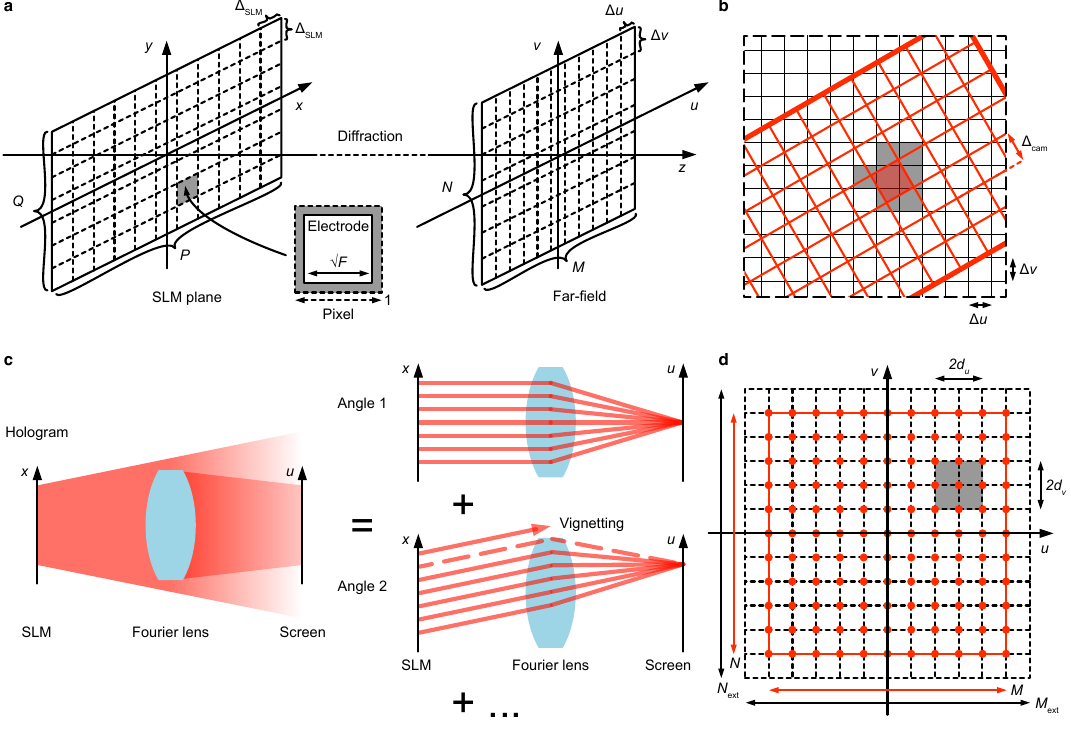}
    \caption{Different aspects of the numerical implementation. (a) Discrete pixel grid, with the pixel structure, and the Fourier-plane sampling grid. (b) Integration of the sampled Fourier-plane field (black grid) by the camera (red grid). For each camera pixel, all Fourier-plane samples (grey) overlapping with the camera pixel (red) must be integrated, weighted by their relative overlap areas. (c) Projection through a lens as a superposition of rays originating at different SLM locations $(x,y)$, directed to different Fourier-plane coordinates $(u,v)$. Depending on the $(x,y)$ origin, and the diffraction angle, the ray may completely miss the lens aperture, giving rise to $A(x,y;u,v)$. (d) For pupil-aberration modelling, PSFs are computed from $\Psi$ on a corner-to-corner node grid (red dots) within the first order (red square), convolved with overlapping far-field tiles (grey), and bilinearly interpolated to reconstruct the final field. To avoid edge effects, the computation extends the first-order sampling $M \times N$ to $M_\text{ext} \times N_\text{ext}$ at this stage, and the extra border is discarded afterwards.}
    \label{fig:si_diagrams}
\end{figure}

Continuous quantities are denoted by italic symbols, and sampled quantities are denoted by bold symbols. The coordinates in the SLM plane are defined in units of the SLM pixel pitch. The SLM contains $P \times Q$ pixels with indices $(p,q)$ and pixel-centre coordinates $(x_p,y_q) = (p-P/2, q-Q/2)$. The pixel aperture consists of a modulating electrode and a surrounding deadspace,
\begin{equation}
\label{eqn:pix_ap}
\begin{split}
    A_\text{pixel}(x,y) &= \Pi_\text{elec}(x,y) + \rho (\Pi_\text{pixel}(x,y) - \Pi_\text{elec}(x,y)) \\
    &=
    \Pi_\text{elec}(x,y) + \rho \Pi_\text{dead}(x,y),
\end{split}
\end{equation}
where $\Pi_\text{pixel}$ is the rectangular function of pitch $1$, $\Pi_\text{elec}$ and $\Pi_\text{dead}$ are rectangular functions that describe the electrode and deadspace regions, respectively, and $\rho$ is the reflectivity of the deadspace. The electrode defines the pixel fill factor $F$, equal to the ratio of the electrode area to the pixel area, such that the electrode side length is $\sqrt{F}$ (Fig.~\ref{fig:si_diagrams}a). The resulting SLM aperture function is
\begin{equation}
    A_\text{SLM}(x,y) = \sum_{p,q} A_\text{pixel}(x-x_p,y-y_q).
\end{equation}
The phase response of the SLM is modelled by a lookup table (LUT) $\phi_\text{LUT}$ relating input greyscale values $\mathbf{g}[p,q] \in [0,1]$ to the phase delays imparted by the corresponding pixels on the incident light,
\begin{equation}
    \bm{\phi}_\text{ideal}[p,q] = \phi_\text{LUT}(\mathbf{g}[p,q]).
\end{equation}
The LUT may deviate from the supposed linear calibration, and in idealised conditions modulates the reflected phase profile just above the electrodes
\begin{equation}
    \phi_\text{ideal}(x,y) = \sum_{p,q} \bm{\phi}_\text{ideal}[p,q] \Pi_\text{elec}(x-x_p,y-y_q).
\end{equation}
Fringing electric fields and LC coupling cause the phase response of each pixel to extend beyond its electrode area. The resulting inter-pixel crosstalk is modelled as a spatial convolution with kernel $\kappa(x,y)$, giving the effective phase modulation
\begin{equation}
\label{eqn:crosstalk}
    \phi(x,y) = \phi_\text{ideal}(x,y) * \kappa(x,y).
\end{equation}
The incident field $\vec{E}_\text{in}$ is incident at polarisation angle $\theta_\text{pol}$ to the SLM's LC director and can be related to the reflected $\vec{E}_\text{out}$ using Jones matrices,
\begin{equation}
\label{eqn:polarisation}
    \vec{E}_\text{out} = 
    \begin{bmatrix}
        e^{j\phi(x,y)} & 0 \\
        0 & 1
    \end{bmatrix}
    \vec{E}_\text{in} 
\end{equation}
where the constant global phase cannot be measured and is neglected. Therefore, only the component of the field aligned with the LC director is modulated. Without an analyser in the system, the perpendicular component has typically low frequencies of the laser mode and forms part of the incoherent \emph{zeroth (diffraction) order} in the formed image. It can be reduced with careful alignment and we assume it is negligible here. For this reason, only the modulated part is studied further and its $\cos{\theta_\text{pol}}$ term can be absorbed into the overall magnitude of the incident field as it cannot be separated in the final model. The reflected optical field can be described in more detail,
\begin{equation}
\label{eqn:slm_field}
    E(x,y) = E_\text{SLM}(x,y) e^{j\phi(x,y)} A_\text{SLM}(x,y).
\end{equation}
where $E_\text{SLM}(x,y)$ is the ``effective'' illumination on the SLM's surface. We call it this way because $E_\text{SLM}(x,y)$ absorbs all perturbations from dust present in the optics and phase ramps from tip/tilt aberrations that can be brought into the SLM plane and merged directly with this term. The propagation to the focal plane of a Fourier lens is modelled using Fraunhofer diffraction \cite{Goodman2017}, with lens aberrations modelled as done in Ref.~\citenum{Wojcik:26}, using four Seidel and tip/tilt coefficients. The aberrations and finite lens aperture are represented by the pupil function
\begin{equation}
    \Psi(x,y;u,v) = A(x,y;u,v) e^{j \Phi(x,y;u,v)},
\end{equation}
where $A$ denotes the aperture function and $\Phi$ the wavefront phase error. In an ideal system, all rays associated with a given focal-plane coordinate $(u,v)$ accumulate identical optical path lengths and are transmitted by the lens aperture. Real lenses deviate from this behaviour due to aberrations and finite aperture size. As a result, the optical path length and transmission of a ray depend on both its position $(x,y)$ in the SLM plane and its propagation direction, represented by the focal-plane coordinates $(u,v)$.

The phase function $\Phi(x,y;u,v)$ models the resulting wavefront error relative to an ideal Fourier-transform lens, while the aperture function $A(x,y;u,v)$ describes whether a given ray is transmitted through the finite lens aperture (Fig.~\ref{fig:si_diagrams}c). Together, these effects are captured by the pupil function $\Psi(x,y;u,v)$. In this work, the lens is placed relatively close to the SLM and is assumed to pass all rays ($A = 1$).

The focal-plane field is
\begin{equation}
    U(u,v) = B(u,v) + \iint_{\mathbb{R}^2} E(x,y) \Psi(x,y;u,v) e^{-2\pi j(ux+vy)} \, \mathrm{d}x \, \mathrm{d}y,
\end{equation}
where $B(u,v)$ is coherent stray light scattered from the optics. The constant terms in front of the integral in the Fraunhofer diffraction formula are neglected here because they only change the global magnitude and phase, which cannot be determined. This magnitude is additionally scaled by an inseparable combination of reflectivities of the electrodes and the polarisation angle $\theta_\text{pol}$, so we merge them in a single normalised constant.

The focal-plane intensity is
\begin{equation}
    I(u,v) = |U(u,v)|^2.
\end{equation}
The coordinates $(u,v)$ are spatial frequencies conjugate to the SLM-plane coordinates $(x,y)$. Since $(x,y)$ are expressed in units of the SLM pixel pitch $\Delta_{\text{SLM}}$, $(u,v)$ have units of the reciprocal pixel pitch ($1 / \Delta_{\text{SLM}}$). Under the paraxial approximation, these are related to the diffraction angles $(\theta_u, \theta_v)$ and the spatial coordinates $(\xi, \eta)$ in the focal plane, by
\begin{equation}
    \xi = f\theta_u = \frac{f\lambda}{\Delta_{\text{SLM}}}u,
    \qquad
    \eta = f\theta_v = \frac{f\lambda}{\Delta_{\text{SLM}}}v,
\end{equation}
where $f$ is the focal length of the Fourier lens and $\lambda$ is the wavelength. The periodic sampling imposed by the SLM pixel grid causes the Fourier plane to repeat periodically in both dimensions. The \emph{first diffraction order} denotes the central unit cell, which is the Nyquist band of the SLM, 
\begin{equation}
    (u,v) \in [-0.5,0.5)^2,
\end{equation}
whose boundaries correspond to diffraction angles $\theta_u,\theta_v \approx \pm \lambda/(2\Delta_{\text{SLM}})$. This square shape with unit side length in $(u,v)$ is caused directly by the square SLM pixels.

The intensity pattern is recorded by a camera composed of $W \times H$ pixels with pitch $\Delta_{\text{cam}}$. Any misalignment between the optical and sensor coordinate systems is modelled by the affine transformation
\begin{equation}
    \mathcal{T}:
    (u',v')^\mathsf{T} \mapsto (u,v)^\mathsf{T}
    = \mathbf{A}(u',v')^\mathsf{T} + \mathbf{t},
\end{equation}
where $\mathbf{A}$ is a $2 \times 2$ matrix (rotation $\theta$ and scaling $\mathbf{s}$) and $\mathbf{t}$ is a translation vector - no shear is allowed. $\mathcal{T}$ defines the camera-frame intensity
\begin{equation}
    I_\text{cam}(u',v')
    =
    I\!\left(\mathcal{T}(u',v')\right)
    =
    I\!\left(u,v\right).
\end{equation}
The saturation signal corresponding to the camera full well capacity is
\begin{equation}
    C_{\text{sat}}
    =
    \frac{N_{\text{sat}} hc}
    {\lambda \eta_{\text{QE}}},
\end{equation}
where $N_{\text{sat}}$ is the full well capacity of the pixel, $\eta_{\text{QE}}$ is the quantum efficiency, $h$ is Planck's constant, and $c$ is the speed of light. The recorded image is then obtained by integrating the incident intensity over the respective camera pixels (Fig.~\ref{fig:si_diagrams}b) and the exposure time $T_\text{exp}$, and clipping the result at $C_{\text{sat}}$:
\begin{equation}
    \mathbf{C}[r,s]
    =
    \min\left(
    \int_0^{T_\text{exp}}
    \iint_{\text{pixel}(r,s)}
    I_{\text{cam}}(u',v')
    \, \mathrm{d}u' \mathrm{d}v' \mathrm{d}\tau,
    C_{\text{sat}}
    \right).
\end{equation}
We use this clipping to define $I_\text{sat} = C_\text{sat} / (\beta \Delta_\text{cam}^2 T_\text{exp})$ where $\beta$ is the unknown global magnitude factor absorbed into the forward model. Thus, $I_\text{sat}$ corresponds to the effective intensity required to saturate the camera. All reported intensities and amplitudes are expressed in units of $I_\text{sat}$.

\subsection*{Discretisation}
The continuous model is discretised for efficient numerical implementation. To accurately model inter-pixel crosstalk, each SLM pixel is subdivided into $K \times K$ square sub-pixels, giving a $PK\times QK$ sampling grid with coordinates
\begin{equation}
    (x'_k,y'_l) =
    \left(
        \frac{k}{K}-\frac{P}{2},
        \frac{l}{K}-\frac{Q}{2}
    \right).
\end{equation}
Throughout this section, the operator $/\!/$ denotes floor (integer) division,
\begin{equation}
    a /\!/ b \equiv \left\lfloor \frac{a}{b} \right\rfloor.
\end{equation}
The ideal phase profile is assumed constant within each pixel and is upsampled to the sub-pixel grid as
\begin{equation}
    \bm{\phi}_{\text{ideal},\uparrow}[k,l] = \bm{\phi}_{\text{ideal}}[k/\!/K,l/\!/K].
\end{equation}
This approximation neglects phase variations within the deadspace, justified by the high fill factor ($F \gtrsim 0.9$) of typical devices. Crosstalk is incorporated by convolving the upsampled phase map with a discretised kernel $\bm{\kappa}$,
\begin{equation}
    \bm{\phi}_{\text{conv},\uparrow}
    = \bm{\phi}_{\text{ideal},\uparrow} * \bm{\kappa}.
\end{equation}
Representing the phase as piecewise constant over the sub-pixel grid,
\begin{equation}
\label{eqn:mod_disc}
    e^{j\phi(x,y)} 
    \approx \sum_{k,l} e^{j\bm{\phi}_{\text{conv},\uparrow}[k,l]} \Pi_{\text{sample}}(x-x'_k,y-y'_l),
\end{equation}
where $\Pi_{\text{sample}}$ is a square aperture of width $1/K$. The incident illumination field is sampled on the SLM pixel grid and upsampled to the sub-pixel grid as
\begin{equation}
    \mathbf{E}_{\mathrm{SLM},\uparrow}[k,l]
    = E_{\mathrm{SLM}}(x_{k//K},y_{l//K}).
\end{equation}
The pixel aperture in $A_\text{SLM}$ is independent of the greyscale input follows the periodicity of the pixel pitch, so it does not need to be sampled at all: its entire effect can be carried in the Fourier domain. Separating the reflected field of Eq.~(\ref{eqn:slm_field}) into the modulated field and the aperture,
\begin{equation}
    E(x,y)
    = \tilde{E}(x,y) A_\text{SLM}(x,y),
\end{equation}
where $\tilde{E}(x,y) = E_\text{SLM}(x,y) \exp({j\phi(x,y)})$ captures all the greyscale dependence and all periodic pixel structure to $A_\text{SLM}$.

Neglecting $\Psi$, if $\tilde{E}$ were constant across each pixel, $\tilde{E}[p,q]$, the two terms would separate exactly under the Fourier transform,
\begin{equation}
\begin{split}
    \mathcal{F}\{\tilde{E}A_\text{SLM}\}(u,v)
    &= \mathcal{F}\left\{ \sum_{p,q} \tilde{E}[p,q] A_\text{pixel}(x-x_p,y-y_q) \right\} \\
    &= \mathcal{F}\left\{ A_\text{pixel}(x,y) * \sum_{p,q} \tilde{E}[p,q] \delta(x-x_p,y-y_q) \right\} \\
    &= \hat{A}_\text{pixel}(u,v) \sum_{p,q} \tilde{E}[p,q] e^{-2\pi j(ux_p+vy_q)},
\end{split}
\end{equation}
using the convolution theorem, where $\hat{A}_\text{pixel} = \mathcal{F}\{A_\text{pixel}\}$, so that the aperture acts purely as a multiplicative envelope in the far field. 

Inter-pixel crosstalk breaks this exact separation by making $\tilde{E}$ vary within a pixel, coupling it to neighbouring diffraction orders. For typical high-fill-factor SLMs, this coupling is weak. Neglecting it lets us maintain the decoupled form as an approximation, but we must return to continuous operator notation to capture sub-pixel spatial variation:
\begin{equation}
    \mathcal{F}\{\tilde{E} A_\text{SLM}\}(u,v) \approx \frac{\hat{A}_\text{pixel}(u,v)}{\mathrm{sinc}(u) \mathrm{sinc}(v)} \mathcal{F}\{\tilde{E}\}(u,v).
\end{equation}
The $\mathrm{sinc}(u) \mathrm{sinc}(v)$ term enters as a decoupling correction: treating $\tilde{E}$ as a continuous, piecewise-constant field inherently embeds a default $100\%$ fill factor sinc envelope into $\mathcal{F}\{\tilde{E}\}$, and dividing by it removes this envelope. This prevents double-counting when applying the physical pixel aperture $\hat{A}_\text{pixel}(u,v)$.

With the physical pixel behaviour isolated, we discretise $\tilde{E}$ on a finer sub-pixel grid to model the crosstalk:
\begin{equation}
    \tilde{E}(x,y) \approx \sum_{k,l} \mathbf{E}[k,l] \Pi_\text{sample}(x-x'_k,y-y'_l), 
\end{equation}
where $\mathbf{E}[k,l] = \mathbf{E}_{\text{SLM},\uparrow}[k,l] \exp{(j\bm{\phi}_{\text{conv},\uparrow}[k,l])}$ represents the upsampled illumination and the crosstalk-modulated phase. The focal-plane field is
\begin{equation}
\label{eqn:semidisc_far}
    U_\text{ideal}(u,v) \approx D(u,v) \sum_{k,l} \mathbf{E}[k,l] e^{-2\pi j(ux'_k+vy'_l)},
\end{equation}
where the combined far-field envelope $D(u,v)$ cleanly balances the sub-pixel sampling cell with the pixel-scale aperture correction:
\begin{equation}
\label{eqn:envelope}
    D(u,v) =
    \left[
    \frac{1}{K^2} \mathrm{sinc}\!\left(\frac{u}{K}\right)\mathrm{sinc}\!\left(\frac{v}{K}\right)
    \right]
    \left[
    \frac{\hat{A}_\text{pixel}(u,v)}{\mathrm{sinc}(u) \mathrm{sinc}(v)}
    \right]
\end{equation}
From Eq.~(\ref{eqn:pix_ap}),
\begin{equation}
\label{eqn:pix_envelope}
    \hat{A}_\text{pixel}(u,v) = (1 - \rho) F \, \mathrm{sinc}\left(u \sqrt{F}\right)\mathrm{sinc}\left(v \sqrt{F}\right)
    +
    \rho \, \mathrm{sinc}\left(u \sqrt{F}\right) \mathrm{sinc}\left(v \sqrt{F}\right).
\end{equation}
Its value at the origin, $\hat{A}_\text{pixel}(0,0)=F+(1-F)\rho$, is degenerate with the global scaling $\beta$, so instead we normalise $D$ to unity at $u=v=0$. The envelope is evaluated over the sampled first diffraction order and its surrounding margin, where $\mathrm{sinc}(u) \mathrm{sinc}(v)$ is bounded away from zero.

Evaluating the Fourier plane in Eq.~(\ref{eqn:semidisc_far}) on the grid $(u_m,v_n)$ reduces the formula to discrete Fourier transform of the $PK\times QK$ array $\mathbf{E}$, which may therefore be computed efficiently using a fast Fourier transform (FFT):
\begin{align}
    \mathbf{U}_\text{ideal}[m,n]
    = \mathbf{D}[m,n] &\sum_{k,l} \mathbf{E}[k,l] e^{-2\pi j(u_m x'_k+v_n y'_l)} 
    = \mathbf{D}[m,n] \mathrm{FFT}\{\mathbf{E}\}[m,n],
    \\
    \text{where} \quad u_m x'_k+v_n y'_l &= \frac{\left( m - \frac{M}{2} \right) \left(k - \frac{PK}{2} \right)}{MK} + \frac{\left( n - \frac{N}{2} \right) \left(k - \frac{QK}{2} \right)}{NK}. \label{eqn:si_phasor}
\end{align}
The term in Eq.~(\ref{eqn:si_phasor}) ensures that the FFT is centred in both domains. In the implementation, the $PK \times QK$ array $\mathbf{E}$ is zero-padded to $MK \times NK$ prior to the FFT, to ensure bandwidth-limited sampling of the Fourier-plane field. The resulting $MK \times NK$ array contains higher diffraction orders ($K \times K$ orders in total), but is then cropped to the first diffraction order, giving the $M \times N$ array $\mathbf{U}_\text{ideal}$.

\begin{figure}[ht]
    \centering
    \includegraphics[width=\textwidth]{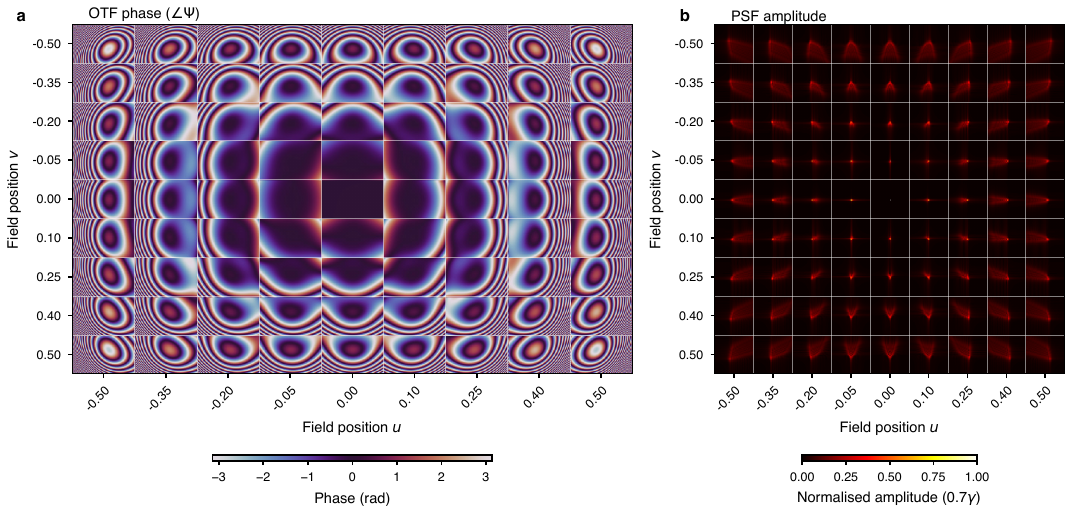}
    \caption{Per-tile pupil aberrations shown with their $(u,v)$-locations for the coefficients measured in the main experiment. (a) The phase of $\Psi$ in the SLM plane for different Fourier-plane tiles. (b) Normalised amplitudes of the corresponding PSFs obtained by Fourier-transforming $\Psi$ shown in (a). They are gamma-corrected for better visibility.}
    \label{fig:si_otf_psf}
\end{figure}

The above treatment neglects the position-dependent aberrations introduced by $\Psi$. A direct computation of with $\Psi$ included is computationally expensive, as it requires a distinct pupil function for every focal-plane coordinate $(u_m,v_n)$. We therefore exploit the weak dependence of $\Psi$ on $(u,v)$ by adopting an isoplanatic patch approximation, extending on the procedure described in Ref.~\citenum{Wojcik:26}. The pupil function is parametrised using four Seidel aberration coefficients together with $x$ and $y$-tilt (tip/tilt) terms.

The method works by defining a regular grid of $N_\text{tile} \times N_\text{tile}$ nodes corner-to-corner over the first diffraction order. At each node, the pupil function $\Psi(x,y;u,v)$ is evaluated and used to compute the corresponding point-spread function (PSF) using an FFT (see Fig.~\ref{fig:si_otf_psf}). The $\mathbf{U}_{\mathrm{ideal}}$ is divided into tiles centred at the nodes (Fig.~\ref{fig:si_diagrams}, as described below, which are then convolved with the corresponding regions in $\mathbf{U}_\text{ideal}$, and then stitched back together to obtain a locally aberrated field $\mathbf{U}_\text{aberr}$. 

Let $d_u = M / N_\text{tile}$ and $d_v = N / N_\text{tile}$ denote the horizontal and vertical node spacings. Each node is associated with a rectangular support of size $2d_u \times 2d_v$ (spanned between the nearest neighbouring nodes), centred on the node itself. As a result, the neighbouring regions overlap by exactly one node spacing in each dimension. The final field is reconstructed by a bilinear blending of the corresponding locally aberrated fields using weights that form a partition of unity, giving a smooth spatially varying approximation of the pupil function without patch-boundary artefacts. This approximation assumes that the PSF is small compared with the $2d_u \times 2d_v$ patch spacing and falls within its area - under sufficiently strong aberration the PSF can extend beyond a patch or move outside, causing the bilinear blend to lose energy and underestimate the true field-dependence of $\Psi$.

The scattered light $B(u,v)$ is discretised to the $M \times N$ array $\mathbf{B}$ on the $(u_m, v_n)$ grid and added to $\mathbf{U}_\text{aberr}$ to give the total Fourier-plane field $\mathbf{U}$, before computing the intensity. In ideal conditions with $2\pi$-range LUT, no crosstalk, flat incident wavefront, full-fill-factor pixels, no scattered light, and no aberrations, the final field reduces to the familiar:
\begin{equation}
    \mathbf{U}[m,n] = \mathrm{sinc}(u_m) \mathrm{sinc}(v_n) \mathrm{FFT}\{e^{j 2 \pi \mathbf{g}}\}[m,n].
\end{equation}

Finally, the integration in the camera model is approximated by first interpolating the intensity over a higher-resolution grid, then affine transforming to the adequately oversampled camera coordinates (by the same integer term), and average-pooling to the camera pixel resolution. The resulting camera-frame intensity is then clipped to the saturation level $C_{\text{sat}}$ as described in the continuous model.

\textbf{Important remarks:} The above formulation assumes that the inter-pixel crosstalk is sufficiently strong and the deadspace sufficiently narrow, such that the phase response is determined solely by the greyscale input. If the deadspace introduced an additional phase delay independent of the greyscale value, the reflected field would contain an additive static contribution originating from the deadspace regions. Since this contribution is independent of the displayed hologram, it would be indistinguishable from the coherent background field $B(u,v)$ and could therefore be absorbed into that term without modifying the overall structure of the model.

Similarly, we assumed that there are no incoherent sources of light in the system (such as the perpendicular polarisation in Eq.~(\ref{eqn:polarisation}). Any such component cannot be measured with the proposed technique and could not be corrected with the displayed holograms.

\section{Supplementary Note 2: Implementation details}

Several implementation choices were made to improve numerical stability, parameter identifiability, and computational efficiency. The LUT $\phi_{\mathrm{LUT}}$ is represented by a set of $N_\text{LUT}$ uniformly spaced nodes, which define non-negative segment slopes in phase, ensuring that the resulting mapping remains monotonic throughout the optimisation. The crosstalk kernel $\bm{\kappa}$ has a limited spatial range, so it is defined over $K_s \times K_s$ samples defined on $K_p \times K_p$ pixels, where $K_s = K_p K$. We chose $K = 3$ to balance precision and computational efficiency and $K_p = 3$ proved to be sufficient to contain the spread of the measured kernels, resulting in $K_s = 9$. To help the optimisation problem, the crosstalk kernel $\bm{\kappa}$ is represented as the sum of the centred Gaussian component parametrised by two orthogonal standard deviations $(\sigma_x, \sigma_y)$, and a learnable residual term defined on the $K_s \times K_s$ samples. This kernel is discretised on the same sub-pixel grid used for phase propagation. To enforce positivity, we take the absolute value of the raw kernel, which is then normalised to ensure an energy-preserving sum of $1$, thereby keeping the average phase response unchanged. Because the residual term is mathematically degenerate with $(\sigma_x, \sigma_y)$, only the combined kernel $\bm{\kappa}$ itself should be interpreted physically.

The incident illumination field $\mathbf{E}_{\mathrm{SLM}}$ is a complex-valued array on the $P \times Q$ SLM pixel grid. The envelope model in Eqs.~(\ref{eqn:envelope}) and (\ref{eqn:pix_envelope}) is parametrised by $F$ and $\rho$. In general $\rho$ can be complex-valued to account for both amplitude and phase differences relative to the active electrode region. However, these degrees of freedom are closely correlated and degenerate. The complex nature of $\rho$ may cause slow phase variations in the Fourier-plane, but this is a weak effect, and we were unable to retrieve the values consistently even in simulations. For this reason, $F$ is fixed to its nominal value provided by the vendor and only a real value $\rho$ is fitted. This serves as an approximation that determines the rough shape of the resulting Fourier-plane envelope $\mathbf{D}$. 

The entire forward model is implemented in Python, using the PyTorch package to handle differentiable tensor operations, allowing gradients with respect to all physical parameters to be computed with automatic differentiation.

The MS-SSIM pyramid weights $w_i$ are generated using 
\begin{equation}
    w_i = e^{-\frac{(i-\mu)^2}{2\sigma^2}},
\end{equation}
with $\mu=0$ and $\sigma=1.5$ to put more significance on the finest-resolution level and progressively decreasing the contributions of coarser scales. The pyramid has $5$ levels and the weights are normalised to a sum of $1$.

\section{Supplementary Note 3: Training procedure}

\subsection*{Initial parameter values and training data}

\begin{figure}[ht]
    \centering
    \includegraphics[width=\textwidth]{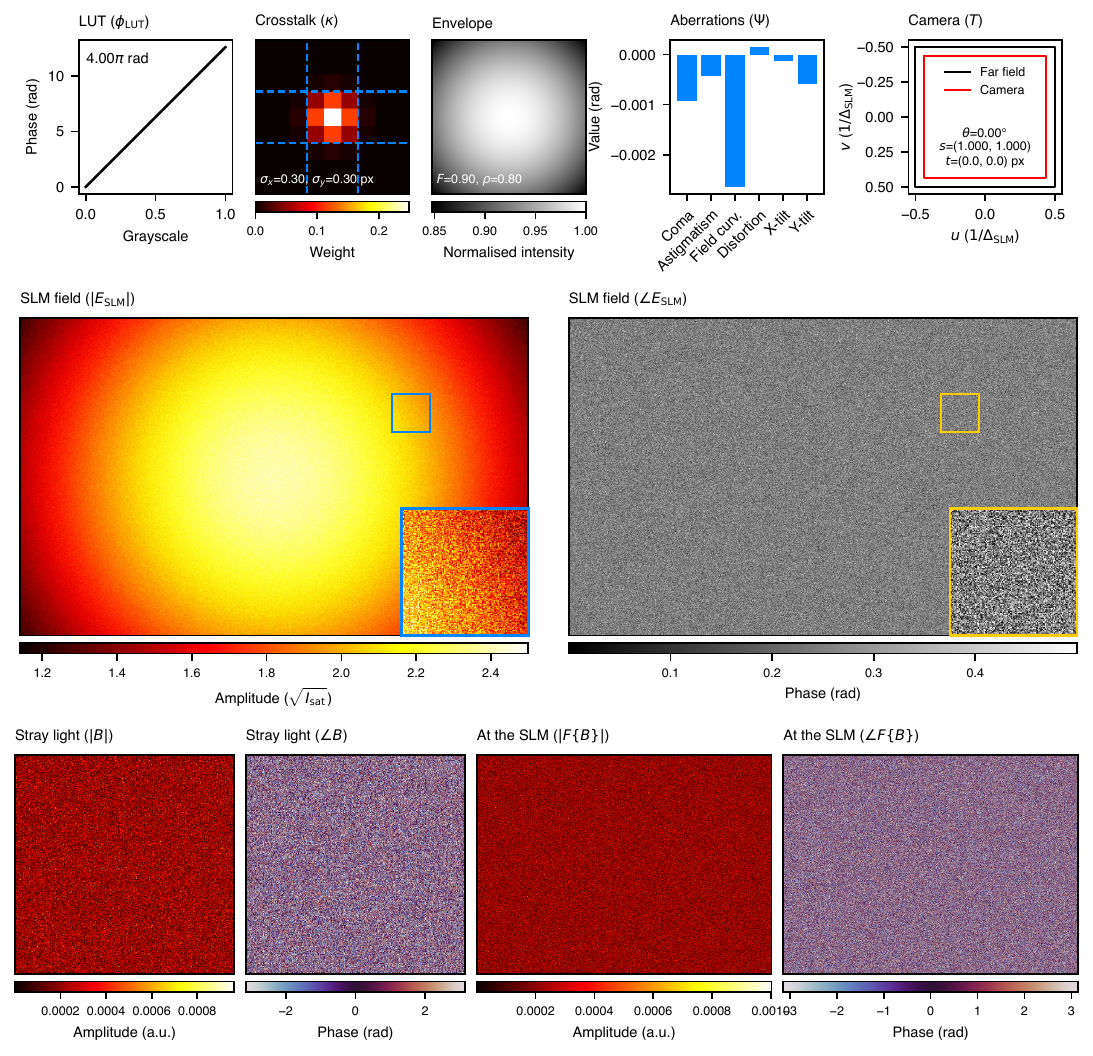}
    \caption{Report of the initial values of each parameter used in the training of the main reported digital twin.}
    \label{fig:si_init_4pi}
\end{figure}

Figure~\ref{fig:si_init_4pi} shows the initial parameter values of the digital twin before training (stage 1 in the main manuscript). Linear nominal phase modulation is assumed in $\phi_\text{LUT}$. To avoid degenerate answers and early stagnation, $\bm{\kappa}$ with $\sigma_x = \sigma_y = 0.3$~pixels, $\rho = 0.8$, the coefficients in $\Psi$ follow a random normal distribution with amplitude of $10^{-3}$, and $\mathbf{E}_\text{SLM}$ has a Gaussian amplitude ($\sigma = 1\times$ the longer SLM side length) with random noise in amplitude and otherwise flat phase have normal noise distribution with magnitudes $0.5$ and $0.1$ respectively. The stray light $\mathbf{B}$ is initialised with a complex-valued normal distribution normalised to the magnitude of $10^{-3}$.

\begin{figure}[ht]
    \centering
    \includegraphics[width=\textwidth]{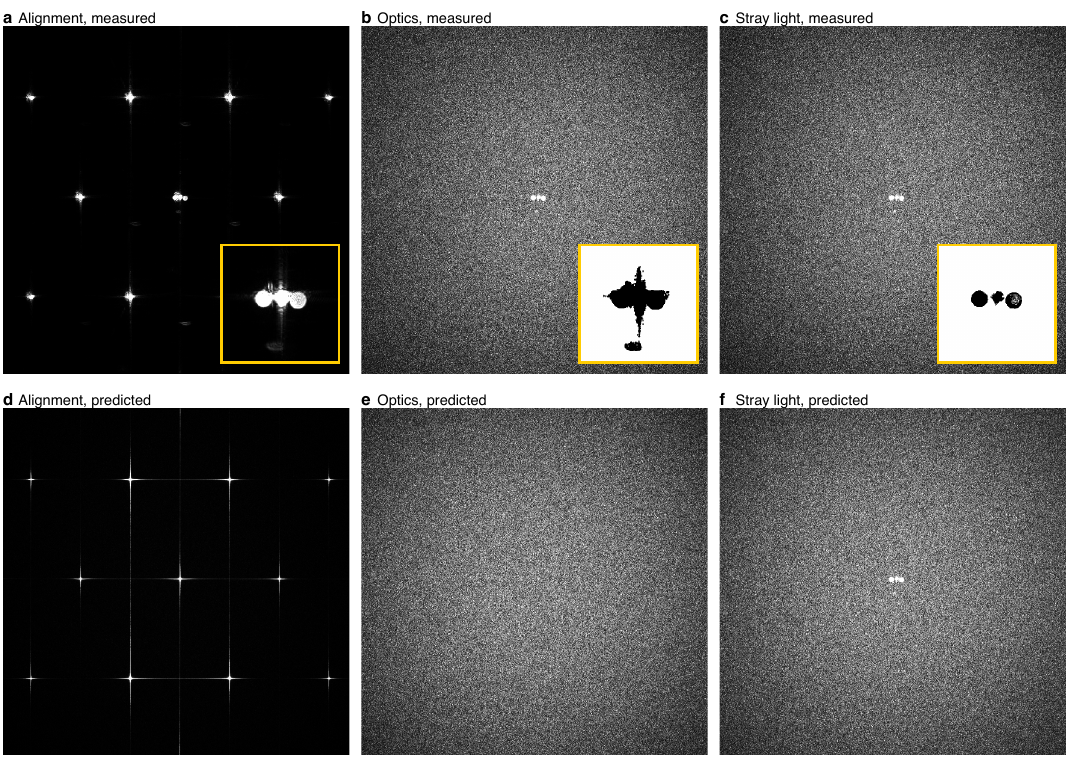}
    \caption{Examples of measured and matched images during training. (a)-(c) Experimental images; (d)-(f) corresponding model predictions. Insets in (a)-(c) show the central portions of the captured zeroth order, the resulting $\mathbf{M}_\text{ZOD}$ mask used for optical identification, and the $\mathbf{M}_\text{sat}$ mask used for stray-light estimation, respectively. (a,d) Chequerboard hologram images used to estimate the affine transform. (b,c) Experimental noise probe images. (e,f) Predicted camera images after optical identification and stray-light estimation, respectively. }
    \label{fig:si_noise}
\end{figure}

Figure~\ref{fig:si_noise} shows sample images from the training dataset, their matched predictions at different stages of optimising the digital twin, and the masks used in the procedure. Panels d-f show how the predictive power of the digital twin improves across the consecutive identification stages.

\subsection*{Learning rates and regularisation}

All gradient-based stages use AdamW with a constant learning rate for the first half of the epochs (warmup), followed by cosine decay to zero. The parameters were split into groups with separate rates, which were determined to balance the speed and stability of convergence (Table~\ref{tab:si_lr}). The pupil coefficients are adjusted at higher rates as they are global and act on large spatial scales, and $\mathcal{T}$  only requires refinement at small rates, since it is estimated by the previous optimisation stage.

\begin{table}[ht]
\centering
\caption{Learning rates of parameter groups used during the training.} 
\label{tab:si_lr}
\begin{tabular}{llll} 
\toprule
Stage & Parameter group & Learning rate & Weight decay \\
\midrule
3 & Optical parameters except $\Psi$ & $5 \times 10^{-2}$ & $10^{-4}$ \\
3 & Pupil $\Psi$ & $2 \times 10^{-1}$ & 0 \\
3 & Camera $\mathcal{T}$ & $5 \times 10^{-3}$ & 0 \\
4 & Stray light $\mathbf{B}$ & $3 \times 10^{-2}$ & $10^{-4}$ \\
\bottomrule
\end{tabular}
\end{table}

Unconstrained, the fields $\mathbf{E}_\text{SLM}$ and $\mathbf{B}$ provide a large number of degrees of freedom and are recovered from the same intensity data, so the same regularisation term $\mathcal{R}$ is applied during their identification. This is done to preserve genuine pixel-sized features such as dust diffraction, while suppressing the noise that would be otherwise overfitted by the optimiser. For a complex field $X$ with amplitude $a = |X|$ and phase $\varphi = \arg{(X)}$,
\begin{equation}
\mathcal{R}(X) = \lambda_{a} \mathrm{TV}(a) + \lambda_{\varphi} \mathrm{TV}_{\text{w}}(\varphi) + \lambda_{\nabla^2} \left\langle (\nabla^2 \mathrm{Re}(X))^2 + (\nabla^2 \mathrm{Im}(X))^2 \right\rangle,
\end{equation}
where $\langle \cdot \rangle$ is the mean over pixels and the weights are given in Table~\ref{tab:si_reg}. The amplitude term uses the isotropic total variation (TV)
\begin{equation}
\mathrm{TV}(a) = \left\langle \sqrt{(\partial_x a)^2 + (\partial_y a)^2 + \epsilon} \right\rangle,
\end{equation}
with $\epsilon = 10^{-8}$ to ensure differentiability at zero gradient. The phase term uses the anisotropic form
\begin{equation}
\mathrm{TV}_{\text{wrap}}(\varphi) = \langle|\partial_x \varphi|_{\text{wrap}}\rangle
+ \langle|\partial_y \varphi|_{\text{wrap}}\rangle,
\end{equation}
where $|\cdot|_{\text{wrap}}$ denotes wrapping the differences to $(-\pi,\pi]$ so that $2\pi$ branch cuts in phase are not penalised as edges. The third term is a discrete Laplacian with Neumann boundaries acting on the real and imaginary parts of the complex field. It penalises pixel-scale roughness in amplitude and phase. All terms are averaged, making the weights independent of grid size.

\begin{table}[ht]
\centering
\caption{Regularisation weights used in $\mathcal{R}$.}
\label{tab:si_reg}
\begin{tabular}{cccccc}
\toprule
Stage & Field & $\lambda_{a}$ & $\lambda_{\varphi}$ & $\lambda_{\nabla^2}$ \\
\midrule
3 & $\mathbf{E}_\text{SLM}$ & $10^{-3}$ & $10^{-3}$ & $10^{-4}$ \\
4 & $\mathcal{F}^{-1}\{\mathbf{B}\}$ & $5\times10^{-2}$    & $0$ & $10^{-4}$ \\
\bottomrule
\end{tabular}
\end{table}

\section{Supplementary Note 4: Alternative system configurations}

\begin{figure}[ht]
    \centering
    \includegraphics[width=\textwidth]{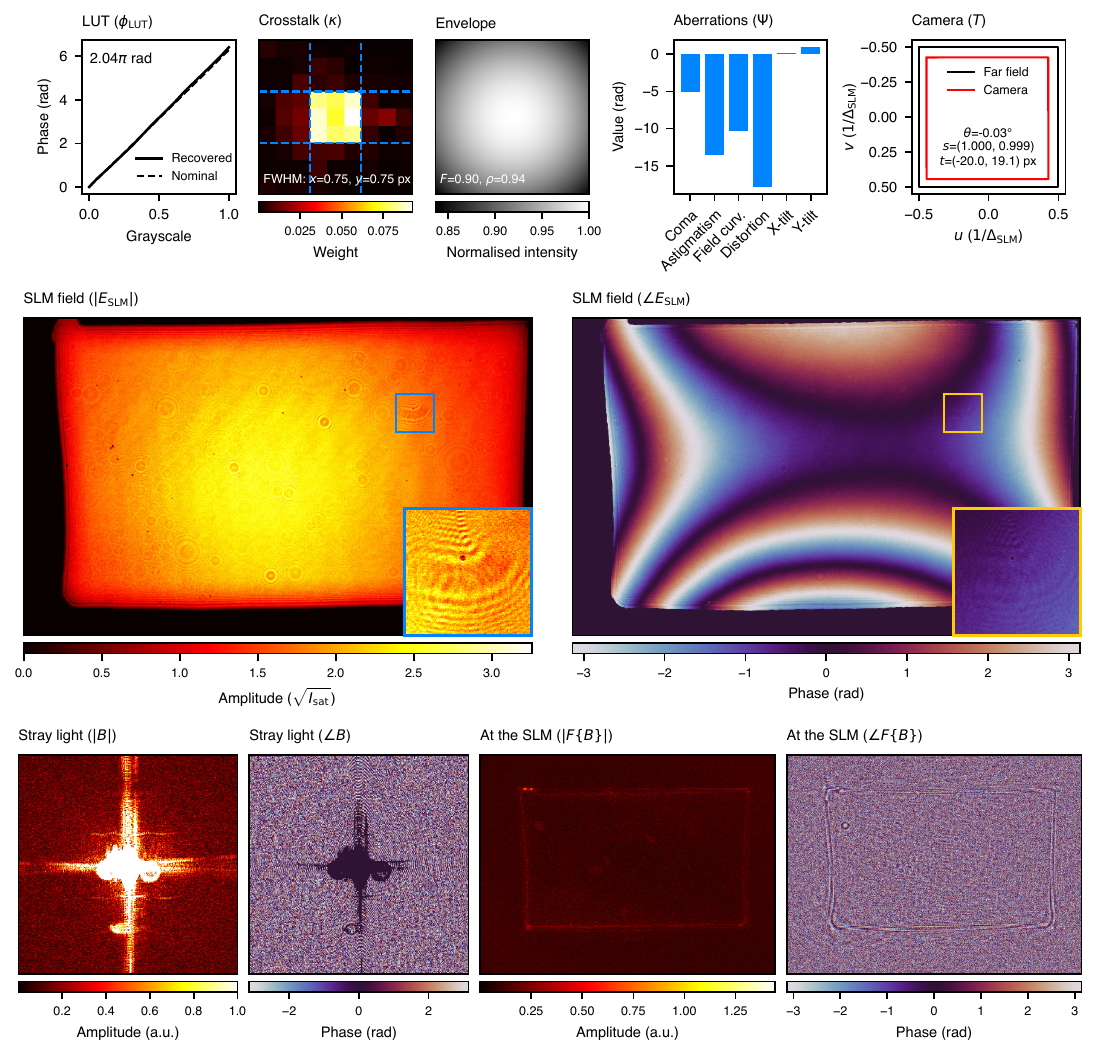}
    \caption{Recovered parameters in the system configured to operate in $2\pi$-phase modulation range, with the aperture placed to reduce stray light.}
    \label{fig:si_recovered_params_2pi}
\end{figure}

\begin{figure}[ht]
    \centering
    \includegraphics[width=\textwidth]{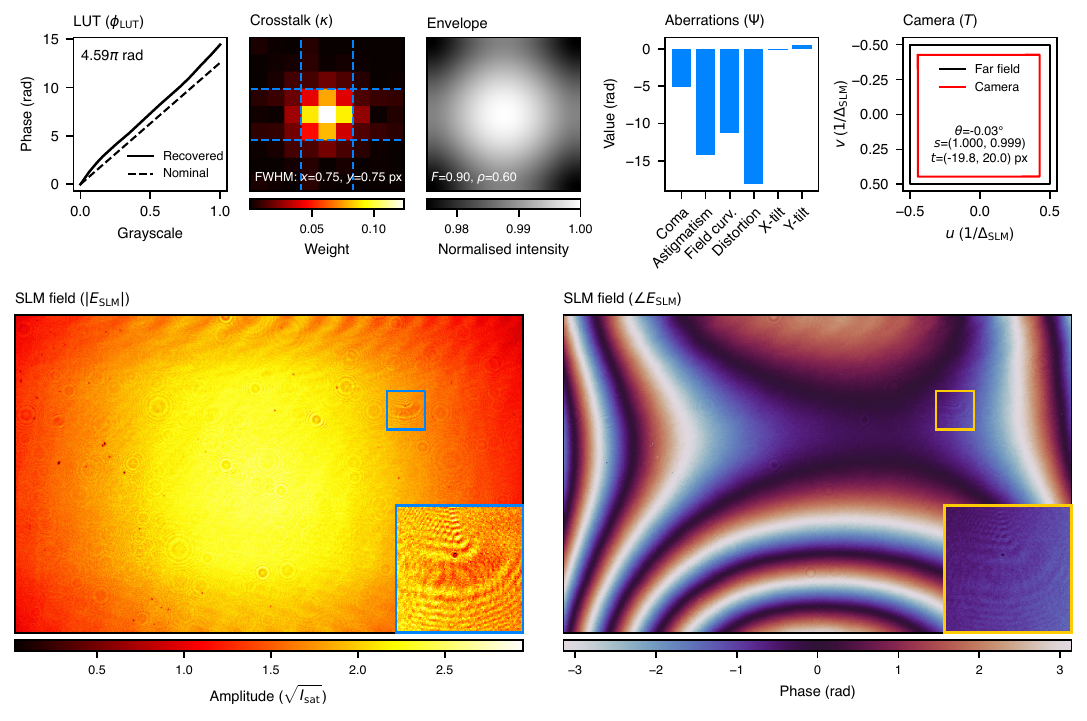}
    \caption{Recovered parameters in the system configured to operate in $4\pi$-phase modulation range, with the aperture removed. The measured stray light $B$ is reported in Figure~6 of the main manuscript.}
    \label{fig:si_recovered_params_4pi}
\end{figure}

Figure~2 of the main manuscript shows the recovered parameter values for the system configuration with $4\pi$ phase modulation range and a 3D-printed aperture inserted to remove most of stray reflections. However, some of the results used alternative experimental configurations:
\begin{enumerate}
    \item $2\pi$ phase modulation, with an aperture.
    \item $4\pi$ phase modulation, without an aperture.
\end{enumerate}
The system was otherwise untouched from its configuration in the main manuscript. The respective reports of the measured parameters are shown in Figures~\ref{fig:si_recovered_params_2pi} and \ref{fig:si_recovered_params_4pi}. 

In the first case, the measured $\phi_\text{LUT}$ aligns closely with its nominal linear behaviour and range, measuring $\SI[parse-numbers=false]{2.04 \pi}{\radian}$. The measured crosstalk is more confined than in the $4\pi$-modulation case - this is likely because of the decreased modulation range, but the exact mechanism depends on the internal operation of the SLM's backplane and the driving board which is not disclosed by the vendor. Under this operation, the SLM seems to generate stronger stray light $B$, closely confined in the centre - this could be due to the deadspace reflections which might have stronger contribution at the lower modulation range. Because of this, the Fourier transform of $B$ shows a more visible shape of the 3D printed aperture and the dust diffraction fringes within it. The pupil aberrations and incident modulation are determined by the optics and remain unchanged.

In the second case, without the aperture, the SLM illumination $\mathbf{E}_\text{SLM}$ reaches all the SLM pixels, which lets the entire SLM area contribute to the projected fields. The shape of recovered $\phi_\text{LUT}$ matches closely in its deviation from the nominal shape with its measurement with the presence of the aperture (Fig.~2 in the main manuscript), also reaching similar recovered $\SI[parse-numbers=false]{4.59 \pi}{\radian}$ range compared to $\SI[parse-numbers=false]{4.57 \pi}{\radian}$. The small deviations in the LUT and the crosstalk parameters may be attributed to the model's assumption of spatial invariance across the SLM surface, which is generally not strictly true in practice. The optimiser converges to a slightly different local optimum under these constraints and the larger active area to fit in this case. However, these minor parametric variations result in a negligible effect on the overall shape and distribution of the projected fields, which remains the definitive metric of the model's performance.

\section{Supplementary Note 5: Training robustness}

Direct experimental assessment of how closely the recovered parameters resemble their true values is difficult, especially because the parameters are coupled and depend on the exact configuration in which the model is used in CGH. The derivation in Supplementary Note~1 assumed various approximations, and because of parameter correlations and degeneracies (for instance, camera location in $\mathcal{T}$ and the phase ramp in $\mathbf{E}_\text{SLM}$), slightly different combinations of values shared between these parameters may either have no significant effect on the formed image or result in the same metric scores. The parameters can be investigated based on the physical features present in the recovered data (such as dust fringes, the presence of the aperture, and consistency amongst different configurations), but the only definite experimental proof that validates the recovery is the resulting improvement in the projected images with CGH. In practice, this is further obscured because the CGH process is not perfect and adds additional noise in the projection.

Simulations, on the other hand, provide a direct way to study the accuracy of the optimisation process. They allow us to create a \emph{ground truth} model with known parameter values, use it to generate synthetic training data, and then compare the recovered parameters directly to this known baseline. This allows us to understand the strengths and limitations of the proposed method. In all tests presented below, the chosen parameters are: $P \times Q = 160 \times 200$ SLM pixels, $\lambda = \SI{500}{\nano\metre}$, $f = \SI{7.5}{\milli\metre}$, $W \times H = 200 \times 200$ camera pixels, $\Delta_\text{SLM} = \Delta_\text{cam} = \SI{4}{\micro\metre}$, unless stated otherwise. This ensures that the camera can observe the formed image. These parameters only rescale the geometry, but the trends observed on these smaller models scale directly to larger pixel counts and more realistic physical parameter values.

\begin{figure}[ht]
    \centering
    \includegraphics[width=\textwidth]{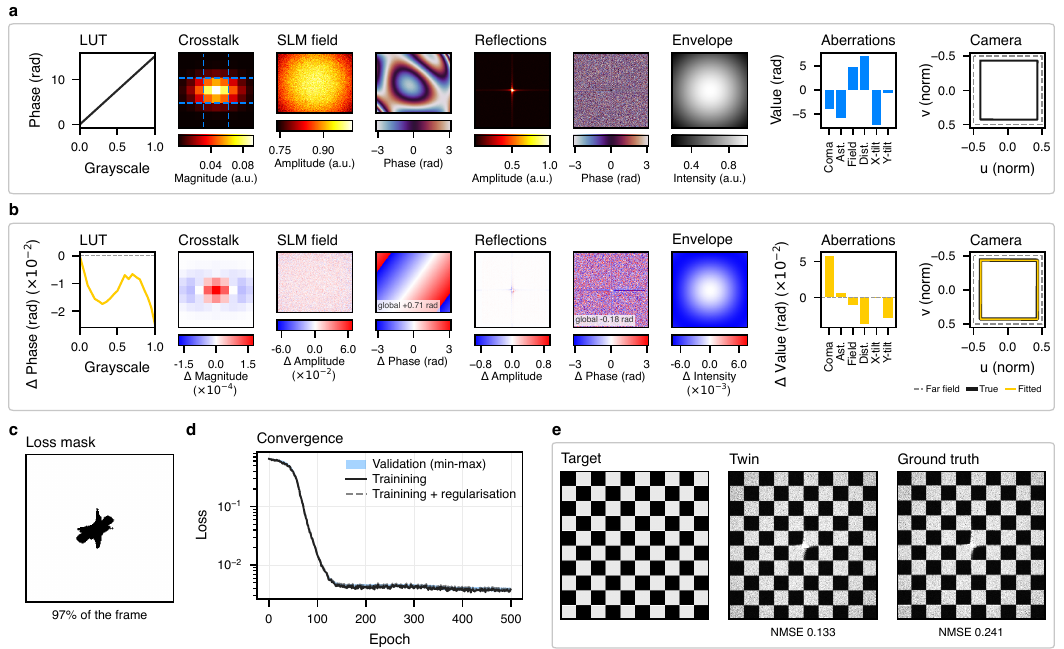}
    \caption{Results of training a digital twin in simulations on synthetic data obtained from a known ground truth reference, with all parameters included in the optimisation. (a) Ground truth parameters. (b) Residual error between the recovered and the ground truth values. (c) Mask $\mathbf{M}_\text{ZOD}$ used in the training of the optical parameter stage (stage~3). (d) Training convergence. (e) Comparison of intensities of example holograms computed on the recovered digital twin, as predicted by the matched twin and displayed in the ground truth system. The NMSE metrics are computed between the projected images and the targets and are shown below each panel; lower is better. }
    \label{fig:si_full}
\end{figure}

Starting with the complete training process on a ground truth with stray light and misaligned camera, the results in Figure~\ref{fig:si_full} show the final loss $\mathcal{L} > 10^{-3}$. This finite loss value is caused by the non-identifiability discussed below, which makes the model stuck in local minima. The single largest mismatch in the recovered twin is the phase ramp in  $\mathbf{E}_\text{SLM}$ relative to the ground truth. It occurs due to the fundamental degeneracy with the shift in $\mathcal{T}$ caused by its Fourier-transform relationship (shift theorem). In general, there is no global anchor with known position in the Fourier plane that would help to avoid this behaviour. Consequently, this biases the translation in the recovered $\mathbf{B}$ field relative to its ground truth counterpart. The trained digital twin is used to generate a batch of $N_B = 4$ holograms while neglecting $\mathbf{B}$. These are subsequently projected with the twin and the ground truth and are frame-averaged. Camera affects are neglected in this comparison, to see how well the underlying intensities match over the entire first diffraction order, which better tests the recovered twin. The comparison between the images results in the NMSE of $0.19$,  most likely due to the translation caused by the shift due to the phase-ramp error.

\begin{figure}[ht]
    \centering
    \includegraphics[width=\textwidth]{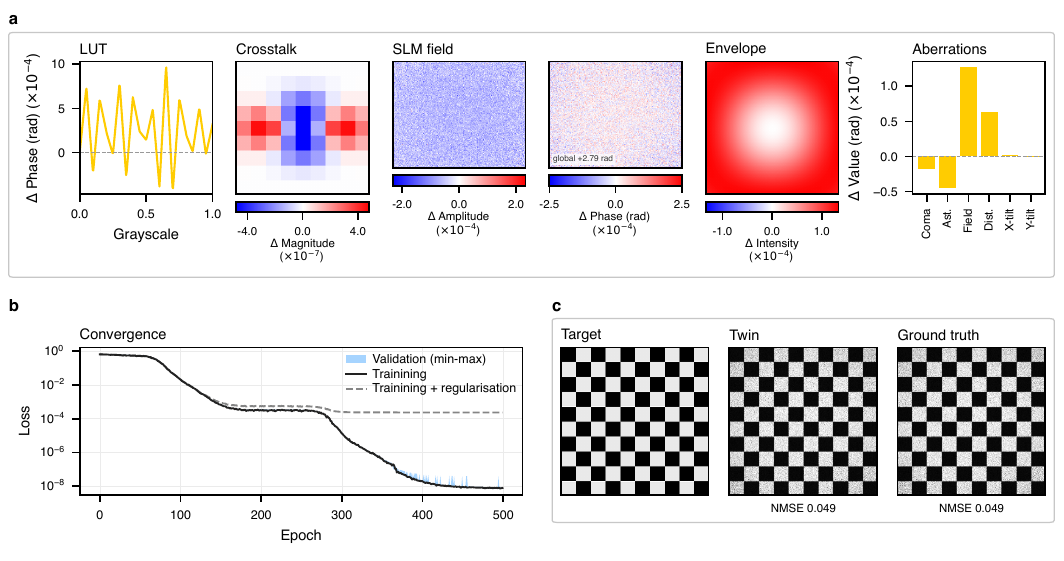}
    \caption{Results of training a digital twin in simulations on synthetic data obtained from a known ground truth shown in Figure~\ref{fig:si_full}. In this case, scattered light is neglected and a known camera alignment is assumed, which significantly improves parameter recovery. (a) Residual error between the recovered and the ground truth values. (b) Convergence plot. (c) Comparison of intensities of example holograms computed on the recovered digital twin, as predicted by the matched twin and displayed in the ground truth system. The NMSE metrics are computed between the projected images and the targets and are shown below each panel; lower is better. }
    \label{fig:si_ideal}
\end{figure}

From this point on, the same ground truth system is used to generate training data but with no scattered light and treating the affine alignment as known exactly and fixed during training to separate these dominant sources of error from the rest of analysis. With these assumptions, stages~2 and 4 of the training procedure can be skipped ($\mathcal{T}$ estimation and $B$ recovery; see Methods in the main manuscript). Figure~\ref{fig:si_ideal} shows the results of the training process. The loss reaches scores below $10^{-8}$ with vanishingly low residual errors in the parameters. The comparison between the frame-averaged intensities predicted with the twin and the one obtained with the ground truth system gives NMSE of $1.19 \times 10^{-8}$, which shows that the model  completely matches the reference.

\begin{figure}
    \centering
    \includegraphics[width=\textwidth]{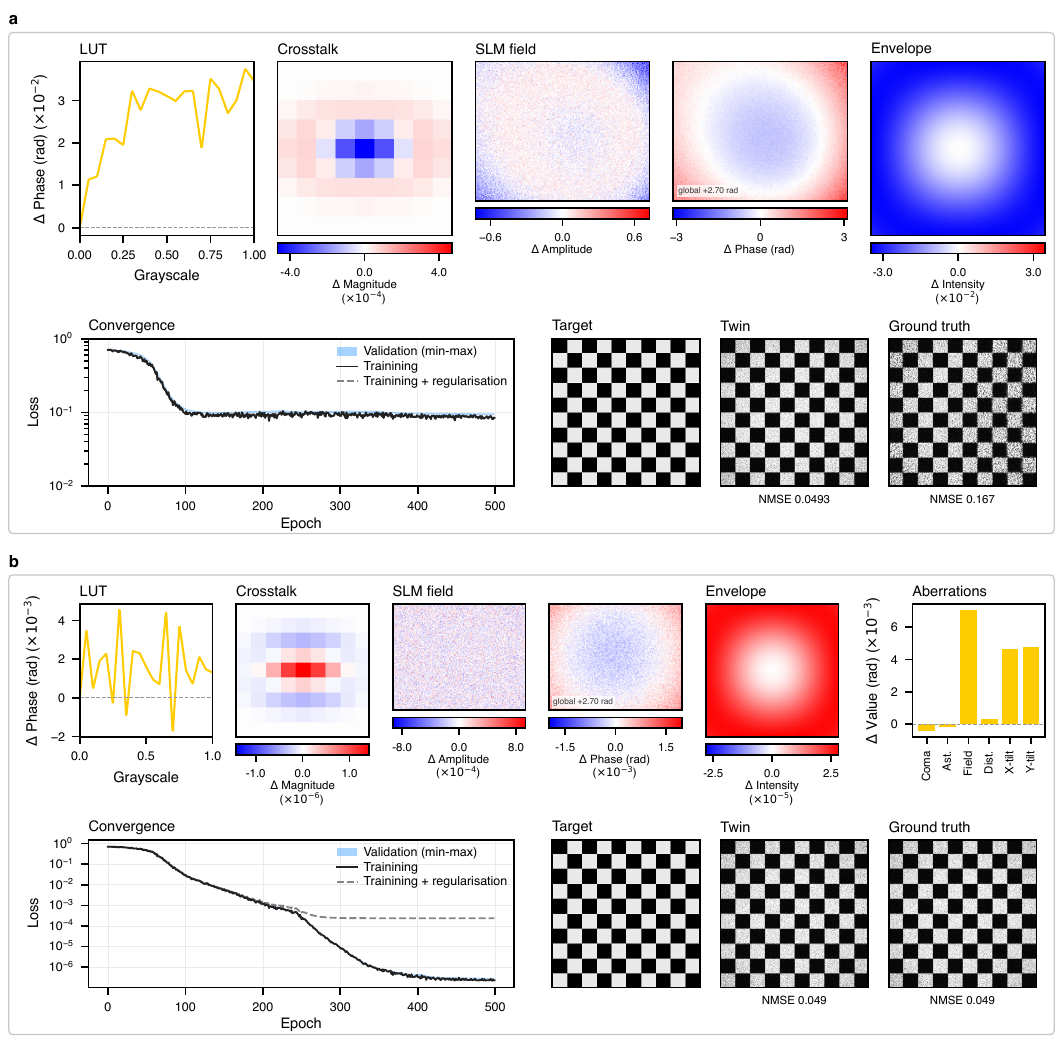}
    \caption{Results of training a digital twin in simulations on synthetic data obtained from a known ground truth shown in Figure~\ref{fig:si_full}, trained only on image data collected from one quarter of the projected field and presented in the same style as Figure~\ref{fig:si_ideal}, with only the discrepancies between the ground truth and the digital twin shown. Two simulated digital twin configurations are tested: (a) neglecting pupil aberrations during training, (b) including pupil aberrations during training. The model (a) which neglects the aberrations fails to generalise outside of the training area, whereas (b) succeeds.}
    \label{fig:si_quarter_field}
\end{figure}

Another important simulation, reported in Figure~\ref{fig:si_quarter_field}, takes a camera that observes only a quadrant of the first diffraction order instead of a larger portion used in other simulations and experiments. When the optimised model does not include pupil aberrations $\Psi$, it can still fit the observed data, but the convergence stops at much higher $\mathcal{L} \approx 0.1$ (Fig.~\ref{fig:si_quarter_field}a). However, this digital twin fails to match the true amplitude and phase of $\mathbf{E}_\text{SLM}$ because this parameter attempts to express both field-invariant $\mathbf{E}_\text{SLM}$ and field-dependent $\Psi$. Most importantly, this distorts the projected images beyond the measured FoV. When $\Psi$ is included in the digital twin and trained, the loss reaches $\mathcal{L} < 10^{-6}$, and the reconstructions generalise over the entire first diffraction order (Fig.~\ref{fig:si_quarter_field}b).

\begin{figure}
    \centering
    \includegraphics[width=\textwidth]{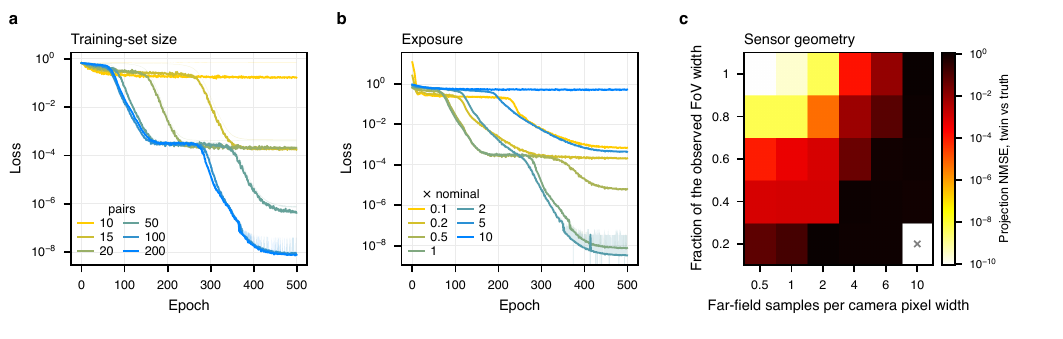}
    \caption{The robustness of the training procedure quantified across different statistics, in simulated training runs. (a) Dependence of convergence on the number of $(\mathbf{g}, \mathbf{C})$ training hologram-image pairs. (b) Dependence of convergence on the relative exposure in the training images $\mathbf{C}$. (c) Dependence on model recovery on the observed fraction of Fourier plane during training and its undersampling by the camera pixels, scored according to the NMSE between the predicted and true projected holograms - lower values indicate closer matching. }
    \label{fig:si_statistics}
\end{figure}

The training method may also be quantified in its resilience to data acquisition conditions. As established earlier, non-zero scattered light $B$ and unknown affine transformation $\mathcal{T}$ degrade the final training metrics, so here we again neglect their effect to determine the limits of the underlying training procedure. Figure~\ref{fig:si_statistics}a shows that the model converges reasonably well, reaching $\mathcal{L} \approx 10^{-4}$ with only $N_S = 15$ training pairs $(\mathbf{g}, \mathbf{C})$, and reaches the numerical error limit within 50-100 samples. In practice, this number is higher because of the lower SNR of real data and the fact that $B$ and $\mathcal{T}$ cannot be neglected. Figure~\ref{fig:si_statistics}b shows training with data ``captured'' at $0.1\times$ to $10\times$ the exposure time relative to the one used in experiments described above. The training converges even in severe underexposure and overexposure, up to $5\times$ the exposure used in other simulations. These results have little meaning as absolute numbers, but rather suggest that the method is robust to large variations in intensity and saturation of the training data. Finally, Figure~\ref{fig:si_statistics}c shows that the method is resilient to severe undersampling between numerical sampling in the Fourier plane and camera pixels ($N_S = 200$ training samples were assumed in all tests). The final recovery depends on how much of the entire FoV is captured by the camera. At full capture, the system can be recovered with up to $6 \times 6$ Fourier-plane samples per pixel, and at full sampling, only $40\%$ of the FoV side length ($16\%$ of the area) needs to be captured for good convergence to $\mathcal{L} \approx 10^{-4}$, or $20\%$ of the FoV side length ($4\%$ of the area) for $\mathcal{L} \approx 10^{-2}$. This is possible because of the Fourier-transform relationship between the SLM and the Fourier plane: due to the Fourier mode mixing, even small regions in the Fourier plane contain information about the entire SLM-plane characteristics.

\newpage

\renewcommand{\refname}{Supplementary References}

\end{document}